\documentclass[journal]{IEEEtran}
\usepackage{graphicx, amssymb, amsthm, amsmath, float, diagbox, caption}
\newcommand{\tabincell}[2]{\begin{tabular}{@{}#1@{}}#2\end{tabular}}

\begin{document}

\title{A New TCP/AQM System Analysis}
\author{Qin~Xu,
        Fan~Li,
        Jinsheng~Sun,
        and~Moshe~Zukerman,~\IEEEmembership{Fellow,~IEEE}
\thanks{Q. Xu  and J. Sun are with the School
of Automation, Nanjing University of Science and Technology, Nanjing, JiangSu Province, P. R. China, 210094
e-mail: missxuqin@hotmail.com}
\thanks{F. Li and M. Zukerman are with Department of Electronic Engineering, City University of Hong Kong, Hong Kong SAR.}}
\maketitle

\begin{abstract}

The MGT fluid model has been used extensively to guide designs of AQM schemes aiming to alleviate adverse effects of Internet congestion. In this paper, we provide a new analysis of a TCP/AQM system that aims to improve the accuracy of the MGT fluid model especially in heavy traffic conditions. The analysis is based on the consideration of two extreme congestion scenarios that leads to the derivation of upper and lower bounds for the queue length and marking probability dynamics and showing that they approach each other in steady state. Both discrete and continuous time models are provided. Simulation results demonstrate that the new model achieves a significantly higher level of accuracy than a simplified version of the MGT fluid model.
\end{abstract}


\section{INTRODUCTION}
\label{sec:Intro}

For over three decades researchers have been seeking efficient and distributed means to control Internet traffic congestion. The common approach has been based on congestion indications generated by links and sent to their sources which then adapt their send rate. The scheme implemented at the links which decides on how and when to indicate congestion is called Active Queue Management (AQM) and the protocol that dictates how to adapt sources rate is mainly TCP (herein TCP means TCP Reno). The combined TCP/AQM system aims to achieve efficient resource utilization, acceptable packet loss and a stable and robust operation.

Congestion indication can be done by either dropping or marking incoming packets. If Explicit Congestion Notifications (ECN) \cite{Floyd:1994:ECN} is enabled, the packets will be marked instead of being discarded, which increases TCP goodput. Given this advantage of ECN, we focus in this paper on the AQM schemes with ECN enabled unless otherwise mentioned.

Many AQM schemes have been proposed. One way to classify AQM schemes is according to their congestion detection method into queue-based, rate-based and queue and rate-based combined \cite{2004:Survey}. RED \cite{1993:Floyd:RED} and PI \cite{Hollot2001PI} are queue-based. Green \cite{2002:Wydrowski:Green}, BLUE \cite{Feng:2002:BLUE} and AVQ \cite{Kunniyur:2001:AVQ} belong to the rate-based class. The combined queue and rate-based category includes REM \cite{Athuraliya:2001:REM}, Yellow \cite{2005:Long:Yellow}, and RaQ \cite{Sun2007RaQ}. An alternative classification into event- and time-driven AQM schemes is based on the method of updating marking probability \cite{Suzer:2012:event}. RED is a typical event-driven AQM scheme. The marking probability of RED is reset at packet arrival events. AVQ and E-AVQ \cite{Wang2012adaptiveAVQ} are also event-driven where resets take place at packet arrivals. BLUE \cite{Feng:2002:BLUE} is event-driven as it updates marking probability when the buffer is empty or full. However, many AQM schemes are time-driven, including PI, PID \cite{Fan:2003:PID}, REM, RaQ. AQM schemes of this type update the marking probability when a certain timeout expires. Table~\ref{table:classificatioin} provides classifications of various AQM schemes.

\begin{table}
\centering
\caption{The Classification of AQM Schemes}
\label{table:classificatioin}
\begin{tabular}{|c|p{2cm}|p{2cm}|}
\hline
\diagbox{Congestion \\ Detection}{Dropping \\Probability \\ Update} & Event-Driven & Time-Driven \\
\hline
Queue Base & RED \cite{1993:Floyd:RED} FPID \cite{2012:chen:FPID}& PI \cite{Hollot2001PI} PID \cite{Fan:2003:PID} MPAQM \cite{2011:Wang:MPAQM} \\
\hline
Rate Based & Green \cite{2002:Wydrowski:Green} BLUE \cite{Feng:2002:BLUE} AVQ \cite{Kunniyur:2001:AVQ} &  \\
\hline
\tabincell{c}{Queue and Rate Based}  & Yellow \cite{2005:Long:Yellow} & REM \cite{Athuraliya:2001:REM} RaQ \cite{Sun2007RaQ}\\
\hline
\end{tabular}
\end{table}

Designing efficient, stable and robust TCP/AQM systems requires the understanding of their dynamics. To achieve such understanding, various models have been proposed \cite{Mathis:1997:model, Kelly:1998:Model, Low:2003:model}. However, the model of \cite{Mathis:1997:model} is based on certain simplified assumptions which introduce significant inaccuracies, the models proposed in \cite{Kelly:1998:Model} and \cite{Low:2003:model} are not scalable. Misra, Gong and Towsley \cite{Misra:2000:model} proposed an analytical model for TCP/AQM systems which is called in \cite{Ajmone2005} the {\it MGT fluid model} after the authors' initials. A simplified version of the MGT fluid model was provided in \cite{Hollot:2001:linearizemodel} and has been widely used. It provided theoretical foundations for numerous analyzes and syntheses of TCP/AQM systems and gave rise to many newly proposed AQM schemes, e.g. \cite{Yu:2011:robust, Wang2008ImcPID}.

In this paper, we use the model of \cite{Hollot:2001:linearizemodel} as a benchmark and we refer to it as the {\it Simplified MGT} model. It considers TCP/AQM as a feedback system where the number of packets allowed to be sent without acknowledgement by the source (sender), designated as the congestion window, is adapted based on congestion indications generated by the AQM (either as lost or marked packets). However, it introduces certain inaccuracy because it assumes independence between the congestion window and the number of lost or marked packets. The inaccuracy tends to increase with increased load. Furthermore, the MGT model of \cite{Misra:2000:model} does not consider practical effects, such as the effect of ECN and modes of TCP operation, (e.g. slow start and congestion avoidance).

In our modeling of a TCP/AQM system we overcome the weaknesses of the Simplified MGT model by considering the above mentioned practical effects. An important distinction between our model and the Simplified MGT model is associated with the roles that TCP and AQM play in the feedback TCP/AQM system. The Simplified MGT model adopts the traditional approach, where TCP is the plant and the AQM is the controller that provides feedback to the TCP. In our model their roles are reversed where the TCP/AQM system is viewed as a system where AQM is the plant and TCP is the controller that provides feedback to the AQM.

We analyze the system dynamics by considering two congestion scenarios which give rise to upper and lower bounds of the queue length and marking probability processes, and show that these bounds are close to each other. Furthermore, we derive a continuous-time model based on differential equations that describe the TCP policy and the link queueing process. We also provide an equivalent discrete-time model for ease of implementation. For a given AQM scheme, our model can provide the transient behavior which include both queue length and marking probability. Both the Simplified MGT model and our model are directly applicable to time-driven AQM schemes and could be applicable to an event-driven AQM if it is approximated by a time-driven system. This paper focuses on time-driven AQM schemes and provides numerical results on PI, REM, RaQ which are all time driven.

The remainder of this paper is organized as follows. In Section~\ref{sec:MGT}, the Simplified MGT model is described. In Section~\ref{sec:Analysis}, we provide an extensive analysis that leads to a new model of a TCP/AQM system which yields bounds for queue length and marking probability dynamics. Section~\ref{sec:Numerical} presents simulation results over a wide range of scenarios and parameter values that validate our new model, compare its accuracy to that of the Simplified MGT model, and illustrate the closeness of the its bounds. Finally, we present our conclusions in Section~\ref{sec:Conclusions}.

\section{THE SIMPLIFIED MGT MODEL}
\label{sec:MGT}

\begin{figure}
\centering
\includegraphics[width=8cm]{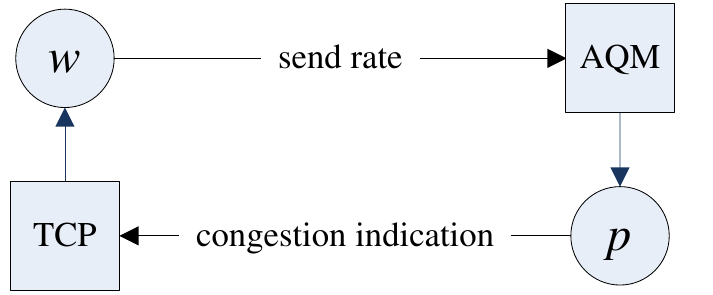}
\caption{Bidirectional relationship of a TCP/AQM system.}
\label{fig:bidirection}
\end{figure}

In a TCP/AQM system, TCP adjusts the send rate to avoid congestion according to the congestion indication received from the AQM. Fig.~\ref{fig:bidirection} shows the working mechanism of a TCP/AQM system. In the figure, the variable $w$ is the congestion window at the TCP end, and $p$ is the marking probability calculated by the AQM.

The original MGT fluid model \cite{Misra:2000:model} assumed that packet losses in a flow follow a non-homogeneous Poisson process. It considers the individual behaviors TCP sessions and their timeout mechanisms. In the Simplified MGT model \cite{Hollot:2001:linearizemodel}, the authors ignored the TCP timeout mechanism and provided a set of differential equations that describe the dynamics of the TCP window at the sources and the queueing at the link for this case. Then, this Simplified MGT model is written as:

\begin{equation}
 \begin{cases}
\ \dot{W}(t)=\dfrac{1}{R(t)} - \dfrac{W(t-R(t))W(t)}{2 R(t)}p(t-R(t))\\
\ \rule{0ex}{2.5em} \dot{q}(t)= -C + N\dfrac{W(t)}{R(t)} ,
 \end{cases}
 \label{eq:modelw}
\end{equation}
where $N$ is the number of TCP sessions, $W(t)$ is the expected congestion window size for a TCP session at time $t\geq0$, $R(t)$ is the expected Round Trip Time (RTT) at time $t\geq0$, $p(t)$ is the marking probability at time $t\geq0$, $C$ is the link capacity, and $q(t)$ is the expected link queue length at time $t\geq0$. The operating point $(W_0, q_0, p_0)$ is defined  by $\dot{W}(t) = 0$ and $\dot{q}(t) = 0$, where $q_0$ denotes the queue length at the operating point and
\begin{equation}
p_0 = \dfrac{2N^2}{R_0^2 C^2},
\label{eq:MGT_p0}
\end{equation}
\begin{equation}
W_0 = \dfrac{R_0 C}{N},
\label{eq:MGT_W0}
\end{equation}
where $R_0$ is the RTT at the operating point.

The Simplified MGT model is known to be accurate for light traffic, but it is inaccurate under heavy traffic. This is partly because the model may yield $p_0 > 1$ in heavy traffic, and then if $p_0$ is truncated to 1, the model may not be able to converge to the target queue size, and if it is not truncated, there the marking probability is erroneous and the queue dynamic may exhibit much slower convergence than the real system (see Section \ref{sec:Numerical}).
Our modeling approach overcomes the $p_0>1$ weakness as discussed below. For brevity, the Simplified MGT model hereafter designates the truncated version of the model. We will also discuss the untruncated version of the Simplified MGT model in Section~\ref{sec:Numerical}.

\section{A NEW TCP/AQM MODEL}
\label{sec:Analysis}

We begin our study of the TCP/AQM system by modeling and analysis of the AQM marking. As mentioned, a time-driven AQM updates its marking probability every sampling interval. This enables the consideration of the sources reactions on the queue length in one sampling period. Next, two scenarios of TCP associated with two extreme congestion levels are considered, which lead to upper and lower bounds for the queue and marking probability dynamics. Both discrete- and continuous-time models are provided. Afterwards, we discuss the effect of ECN on the queue dynamics, followed by an analysis of the operating point where the two scenarios converge at the steady state under the assumption that the entire system is stable. To be able to test the accuracy of the model for a wide range of cases, three non-overlapping and exhaustive congestion levels are then defined and corresponding working scenario(s) are presented. This follows by a discussion on the closeness of the bounds, the system stability and on the setting of a key parameter.

\subsection{Analysis of an AQM Marking Process}
\label{sub:AQM}
In most AQM schemes, packets will be marked or dropped according to marking probability and traffic conditions during a sampling interval. In our analysis, we assume that ECN is always ON and the buffers are sufficiently large so that packet dropping is excluded. Then for the marking process during a sample interval, we make the following assumptions:

\begin{enumerate}
  \item For each packet, there are two mutually exclusive outcomes: marked or unmarked.
  \item During each sampling internal, the marking of each packet is independent of the marking of other packets.
  \item The marking probability remains constant within the duration of a sampling interval.
\end{enumerate}
Accordingly, given the marking probability and the number of packets that arrive during a sampling interval, the number of marked packets within this interval is a Binomial random variable.

Let $m(t)$ be the number of packets that arrive at the link during the time period from $t$ to $t+ \tau$, where $\tau>0$ is an infinitesimally small time period. Let $p(t)$ be the marking probability. As the number of marked packets is Binomially distributed, its expectation is $m(t)p(t)$, and the expected number of unmarked packets is $m(t)(1-p(t))$. Let the sum of all TCP congestion windows $W_s(t)$  increases by $a(t)$ for every unmarked packet, and decreases by $b(t)$ for every marked packet, at time $t\geq0$. Let $\Delta W_s(t+R(t))$ be defined by:
\begin{equation}
\Delta W_s(t+R(t)) \doteq W_s(t+R(t)+\tau)-W_s(t+R(t)) .
\label{eq:delta_W_s}
\end{equation}
Then, we obtain
\begin{equation}
\Delta W_s(t+R(t)) = a(t)m(t)(1-p(t)) + b(t) m(t)p(t).
\label{eq:binomial}
\end{equation}
The function $R(t)$ here is equivalent to the $R(t)$ of the Simplified MGT model used in (\ref{eq:modelw}). It represents the average over all TCP sessions, that are active during the time period from $t$ to $t+\tau$, of their time delay from the moment a packet arrives at a link until its acknowledgement arrives at the sender. This delay is the RTT minus the time it takes the packet to reach the link from the moment it leaves the sender. However, for tractability, $R(t)$ is approximated as the average RTT which is given by
\begin{equation}
\label{eq:RTT}
R(t) = T_p + \dfrac{q(t)}{C},
\end{equation}
where $T_p$ is the average propagation time, $C$ is the link capacity, $q(t)$ is the average queue length during the time period from $t$ to $t+\tau$, and $q(t)/C$ is the mean queueing delay of a packet arriving in that time period. By the approximation we made in (\ref{eq:RTT}), our $R(t)$ is the same as that of the MGT Fluid model which is also based on this approximation.

We assume that $m(t)$ takes the form:
\begin{equation}
m(t) = \dfrac{W_s(t)}{R(t)}  \tau ,
\label{eq:m(t)}
\end{equation}
where ${W_s(t)}/{R(t)}$ represents the packet arrival rate.

Marked packets may belong to different TCP sessions and therefore have different effects on TCP window sizes, for simplicity, we assume that the processes $a(t)$ and $b(t)$ are the same for all sessions. Appropriate values for them will be their expected values over all sessions which will be obtained through the analysis of TCP dynamics in the next subsection.

\subsection{Two Working Scenarios of TCP Data Transmission}
\label{sub:TCP}

\begin{figure}
\centering
\includegraphics[width=0.5\textwidth]{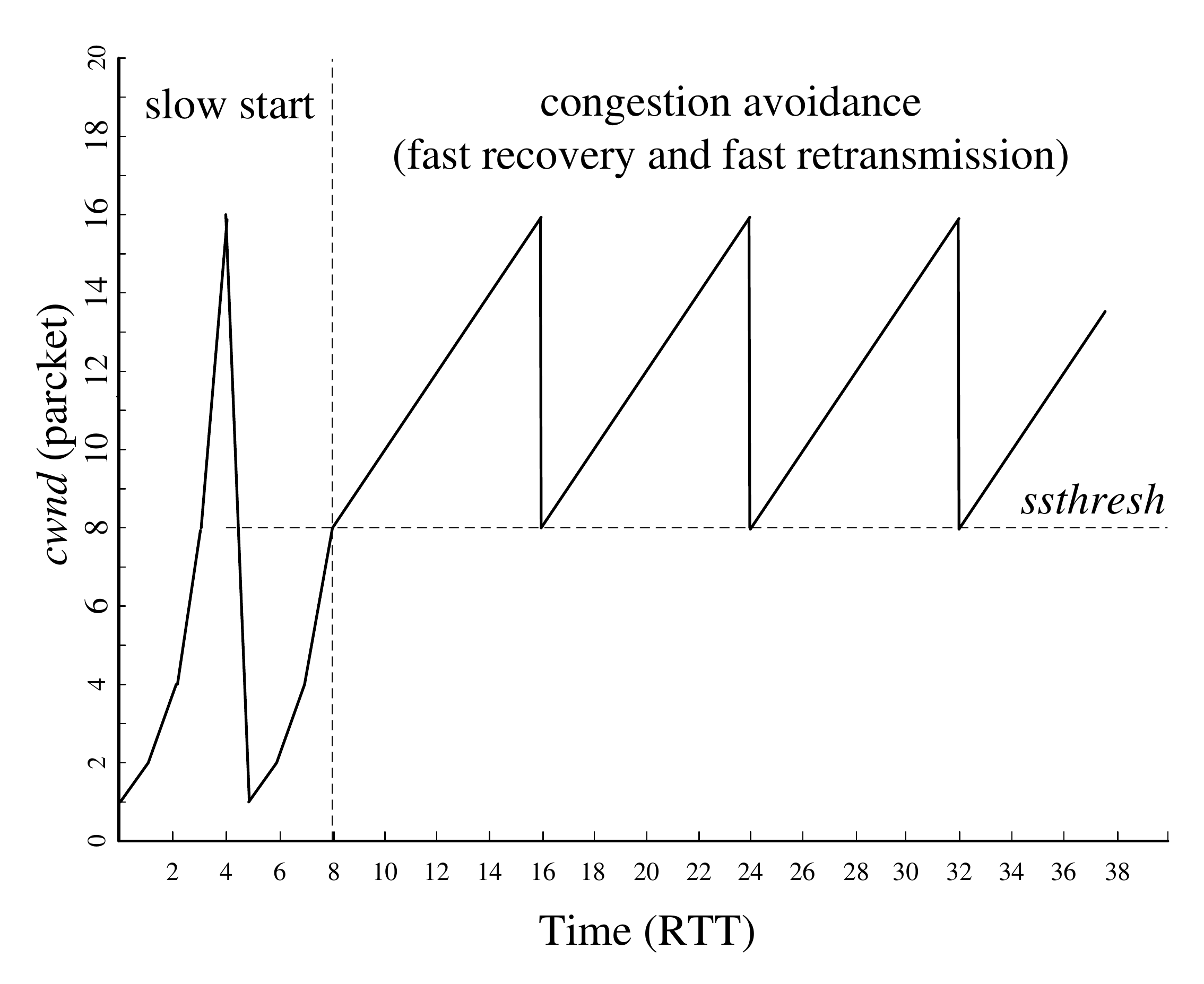}
\caption{A TCP cwnd process.}
\label{fig:four phases}
\end{figure}

TCP is composed of four algorithms \cite{1997RFC2001, 1999RFC2581, 2009RFC5681} which give rise to four data transfer phases: slow start, congestion avoidance, fast retransmit and fast recovery. There are two important variables for these algorithms: congestion window (\emph{cwnd}) that limits the amount of data that TCP can send, and slow start threshold size (\emph{ssthresh}) that determines whether the slow start or congestion avoidance algorithm is used. Slow start (involving exponential growth of \emph{cwnd}) continues until \emph{cwnd} reaches the threshold \emph{ssthresh} which has a lower bound of two packets. Then, congestion avoidance (involving linear \emph{cwnd} growth) is used. When congestion occurs, fast retransmit and fast recovery are used.  A typical single-source single-link TCP cwnd process is illustrated in Fig.~\ref{fig:four phases}.

To introduce our new model, it is convenient to consider the following two scenarios:

\begin{itemize}
 \item Scenario A: In all the TCP sessions \emph{cwnd} $<$ \emph{ssthresh}, so they are either in slow start, or in fast recovery.
  \item Scenario B: In all the TCP sessions \emph{cwnd} $\geq$ \emph{ssthresh}, so they are either in congestion avoidance, or in fast recovery.
\end{itemize}

Scenarios A and B represent two extreme cases where all the TCP sessions behave in a similar way. For tractability, these scenarios exclude cases where in some sessions \emph{cwnd} $<$ \emph{ssthresh} and in others \emph{cwnd} $\geq$ \emph{ssthresh}. Nevertheless, we demonstrate by simulations in Section~\ref{sec:Numerical} that our analysis of  Scenario A and B is sufficient to lead to accurate evaluations for a wide range of system states even if the conditions of neither Scenario A nor B are satisfied.

In the example presented in Fig.~\ref{fig:four phases}, Scenario A takes place at time 0, and then Scenario B takes over at time 8. This example represents a light traffic case involving a single TCP session. As the number of sessions increases, and more sessions need to share a limited capacity, their \emph{cwnds} may be less than 2 packets and therefore less than \emph{ssthresh}, then Scenario A dominates.

To derive the mathematical model of the two scenarios, consider a system where there is a single congested link with capacity $C$. Let $N$ be the number of TCP flows that use this link; they are labeled by $i = 1,  2, \ldots, N$. Then let $W_i(t)$ and $R_i(t)$ denote the window size and the RTT of flow TCP$_i$ ($i = 1, 2, \ldots, N$) at time $t>0$, respectively. In the slow start phase, when an acknowledgement is received by the sender of an unmarked packet, the congestion window will increase by one packet. We obtain:
\begin{equation}
a(t) = \sum_{i = 1}^{N} {\dfrac{W_i(t)}{W_s(t )}} = 1 ,   \qquad  t>0,
\end{equation}
where the ratio ${W_i(t)}/{W_s(t)}$ represents the proportion of total data packets that generated by TCP$_i$.

In the congestion avoidance phase, the congestion window will increase by one \emph{packet} in every RTT, then the increment of the congestion window can be expressed as:
\begin{equation}
\label{aca}
a(t) = \sum_{i = 1}^{N} {\frac{W_i(t)}{W_s(t)} \frac{1}{W_i (t+R(t))}} \approx  \frac{N}{W_s(t+R(t))} , \qquad t>0.
\end{equation}
Notice that in (\ref{aca}) we make the following approximation:
\begin{equation}
\sum_{i = 1}^{N} \frac{W_i(t)}{W_s(t)} \frac{1}{W_i(t+R(t))} \approx \sum_{i = 1}^{N} \frac{W_i(t+R(t))}{W_s(t+R(t))} \frac{1}{W_i(t+R(t))} .
\end{equation}
This approximation is made to simplify the derivation of the window increment $a(t)$, which has negligible affect as packets marking happens repeatedly and $R(t)$ is relatively small. This approximation is also applicable to the derivation of the window decrement $b(t)$.

In the fast recovery process, the congestion window will decrease by half its size when the sender receives a notification of a marked packet. We have
\begin{equation}
b(t) = -\sum_{i = 1}^{N} {\frac{W_i(t)}{W_s(t)} \frac{W_i(t+R(t))}{2}} \approx -\sum_{i = 1}^{N} {\frac{W_i^2(t+R(t))}{2 W_s(t+R(t))}} ,
\end{equation}
where
\begin{align}
\sum_{i=1}^{N} {\frac{W_i^2(t+R(t))}{W_s(t+R(t))}}\geq \frac{(\sum_{i=1}^{N} {W_i(t+R(t))})^2}{W_s(t+R(t))N}, \nonumber \\
\frac{(\sum_{i=1}^{N} {W_i(t+R(t))})^2}{W_s(t+R(t))N}=\frac{W_s(t+R(t))}{N} .
\label{eq:ro}
\end{align}
Then, we write
\begin{equation}
b(t) = -\frac{\rho W_s(t+R(t))}{2N} ,
\end{equation}
where $\rho$ is a function of  $N$ and the $cwnd$ values of all the sessions, and it  satisfies $1 \leq \rho \leq N$.  For large $N$, the extreme cases of $\rho = 1$ and $\rho = N$ are very rare. Having $\rho=1$ corresponds to the case where the congestion windows $W_i (t+R(t))$s of all TCP sessions are the same and nonzero, and $\rho=N$ corresponds to the case where there is always exactly one TCP session with $\emph{cwnd}>0$ and in all other sessions $cwnd = 0$. In practice, both $N$ and the $cwnd$ values of each session vary, so $\rho$ also varies in time. Nevertheless, we will show by simulations in Section~\ref{sec:Numerical} that if $N$ is fixed and if a fixed value for $\rho$ is correctly chosen, the analysis based on a fixed $\rho$ leads to accurate results despite the fact that $cwnd$ values vary. We also show that to model a case where $N$ varies, if we can select the right $\rho$ value for each value of $N$, we still obtain accurate results. Obtaining the parameter $\rho$ is a key issue in our modeling, and although we do not have an analytical way to derive it, we argue and demonstrate in the sequel that setting $\rho = 1$ improves on the accuracy of the Simplified MGT model.

\subsection{Modeling Window Dynamics for Scenarios A and B}
\label{subsec:new_model}

Based on (\ref{eq:binomial}), (\ref{eq:m(t)}) and the above results for $a(t)$ and $b(t)$,  we have the following equation for Scenario A:
\begin{align}
\Delta W_s(t+R(t))=&\dfrac{W_s(t)}{R(t)}(1-p(t))\tau \nonumber \\
&-\dfrac{\rho W_s(t+R(t))W_s(t)}{2N R(t) }p(t) \tau .
\label{eq:SceA}
\end{align}
\setlength{\arraycolsep}{5pt}
Similarly, for Scenario B, we obtain:
\begin{align}
\Delta W_s(t+R(t)) =&\dfrac{N W_s(t)}{R(t) W_s(t+R(t))}(1-p(t))\tau  \nonumber \\
&-\dfrac{\rho W_s(t+R(t))W_s(t)}{2N R(t) }p(t) \tau .
\label{eq:SceB}
\end{align}

Based on above two equations which give the sum of all TCP congestion windows for the two Scenarios, we can present our new models in both discrete- and continuous-time as described in the following.

\subsubsection{Discrete-time Model}
Let time be divided into consecutive fixed-length time intervals each of size $\Delta t$. Let $k$ be an index for these intervals so that the $k$th interval represents the time between $(k-1)\Delta t$ and $k\Delta t$. Recalling our definition of $R(t)$ associated with the TCP sessions that are active between time $t$ and time $t+\Delta t$, let $R(k)$ be the expected RTT of the TCP sessions that are active during the $k$th time interval. Let $n(k)$ be defined as $n(k)=\lfloor R(k)/\Delta t \rfloor$, and $\Delta W_s(k+n(k))$ be defined as $\Delta W_s(k+n(k)) \doteq W_s(k+n(k)+1)-W_s(k+n(k))$. We then obtain the following for Scenario A.
\begin{align}
\Delta W_s(k+n(k)) =& \frac{W_s(k)}{R(k)}(1-p(k)) \Delta t \nonumber \\
&- \frac{\rho W_s(k+n(k))W_s(k)}{2N R(k) }p(k) \Delta t .
\end{align}
For Scenario B, we have:
\begin{align}
\Delta W_s(k+n(k)) =& \frac{N W_s(k)}{R(k) W_s(k+n(k))}(1-p(k)) \Delta t \nonumber \\
&- \frac{\rho W_s(k+n(k))W_s(k)}{2NR(k) }p(k) \Delta t .
\end{align}
In the above two equations, the variables $W_s(k)$ and $p(k)$ are the expected values of the sum of all TCP congestion windows and marking probability at the $k$th time interval, respectively.
\subsubsection{Continuous-time Model}
According to the definition of time-derivative, we have:
\begin{equation}
\dot{W_s}(t+R(t)) = \lim_{dt\rightarrow 0} \frac{W_s(t+R(t)+dt) - W_s(t+R(t))}{dt} .
\end{equation}
Recall the (\ref{eq:SceA}) and (\ref{eq:SceB}), we obtain the following form for Scenario A:
\begin{equation}
\dot{W_s}(t+R(t)) = \frac{W_s(t)}{R(t)}(1-p(t)) - \frac{\rho W_s(t+R(t))W_s(t)}{2N R(t)}p(t) .
\label{eq:SceAcon}
\end{equation}
For the Scenario B, we have:
\begin{align}
\dot{W_s}(t+R(t)) =& \frac{N W_s(t)}{R(t) W_s(t+R(t))}(1-p(t)) \nonumber \\
&- \frac{\rho W_s(t+R(t))W_s(t)}{2N R(t)}p(t) .
\label{eq:SceBcon}
\end{align}

The marking probability $p(k)$ or $p(t)$ depends on the specific AQM scheme, namely, different AQM schemes will update such marking probability based on different measures, for example, PI uses queue length error while REM uses both queue length error and sending rate error. Later in Section \ref{sec:Numerical}, we demonstrate the effect of marking probability of different AQM schemes on the dynamics of queue length and marking probability. As in our model, Scenarios A and B are used as upper and lower bounds, respectively, for the predicted values of queue lengths and marking probability, it is important to have these bounds close to each other. In Subsection~\ref{sub:Closeness} we will provide intuitive arguments and a rigorous proof for the closeness of the bounds, and in Section~\ref{sec:Numerical} we will provide a wide range of numerical results to provide further evidence this closeness.

\subsection{Effect of ECN on the Queue Dynamics}
\label{sub:ECN}
ECN is enabled in some TCP sessions and it is not enabled in others. We will discuss now the different queue dynamics considering operation alternatives when the ECN is ON and OFF. The queue dynamics depends on the packet arrival rate which equals the sum of the send rate of the sources and the packet departure rate which is the link capacity $C$. In the discrete-time model, the queue dynamics is described as:
\begin{equation}
\Delta q(k) = (W_s(k)/R(k) - C) \Delta t ,
\label{eq:DeltaQ_Dis_ECN}
\end{equation}
where $\Delta q(k) \doteq q(k+1)-q(k)$, and $W_s(k)/R(k)$ represents the packet arrival rate.

Equivalently, for the continuous-time model, we obtain,
\begin{equation}
\dot{q}(t) = W_s(t)/R(t) - C ,
\label{eq:DeltaQ_Con_ECN}
\end{equation}
where $\dot{q}(t)$ is the time-derivative of $q(t)$. The first and second term represent the packet arrival and departure rates, respectively.

When the ECN is off, the method of informing the sources of congestion is by dropping packets, instead of marking packets. The queue dynamics, which considers the dropped packets by disabling the ECN, can be expressed in both the discrete- and continuous-time models as:
\begin{equation}
\Delta q(k) = (W_s(k)/R(k) - C - p(k)W_s(k)/R(k)) \Delta t ,
\label{eq:DeltaQ_Dis_NECN}
\end{equation}
and
\begin{equation}
\dot{q}(t) = W_s(t)/R(t) - C - p(t)W_s(t)/R(t) ,
\label{eq:DeltaQ_Con_NECN}
\end{equation}
where the third term corresponds to the dropped packets, which is often ignored in the literature. However, it is important since the operating point can be affected by the dropped packets. This will be discussed next.

\subsection{The Operating Point}
\label{sub:Opera}

The operating point ($W_{s0}$, $q_0$, $p_0$) is defined by $\dot{W_s}(t)=0$ and $\dot{q}(t)=0$ according to the continuous-time model, so that
\begin{equation}
\label{eq:w0}
\dot{q}(t)=0 \Rightarrow W_{s0} = \left\{
    \begin{array}{rcl}
    R_0C ,      &      & \text{if ECN is ON}\\
    \rule{0ex}{2em} \dfrac{R_0C}{1-p_0} ,      &      & \text{if ECN is OFF}
    \end{array} \right.
\end{equation}
\begin{equation}
\label{eq:RTT_op}
R_0 = T_{p} + \dfrac{q_0}{C} ,
\end{equation}
where $W_{s0}$, $q_0$ and $p_0$ are the sum of the congestion windows, the queue length, and the marking (dropping) probability when TCP/AQM system is in steady state, respectively. The value of $W_{s0}$ as shown in (\ref{eq:w0}) is dependent on whether ECN is ON or OFF.

In order to obtain $p_0$ for Scenario A, we write
\begin{equation}
\dot{W_s}(t)=0 \Rightarrow p_0 = \dfrac{1}{1+\frac{\rho W_{s0}}{2N}} = \dfrac{2N}{2N + \rho W_{s0}} .
\label{eq:p0a}
\end{equation}
For Scenario B, we obtain
\begin{equation}
\dot{W_s}(t)=0 \Rightarrow p_0 = \dfrac{1}{1+\frac{\rho W_{s0}^2}{2N^2}} = \dfrac{2N^2}{2N^2 + \rho W_{s0}^{2}}.
\label{eq:p0b}
\end{equation}
Note that the value of $p_0$ in (\ref{eq:p0a}) and (\ref{eq:p0b}) cannot exceed 1, which is an important property that helps improve accuracy over the Simplified MGT model where $p_0$ in (\ref{eq:MGT_p0}) can take values higher than 1.
Specifically, equation (\ref{eq:p0a}) and (\ref{eq:p0b}) correct such error by adding the value $\rho W_{s0}$ or $\rho W_{s0}^2$ to the denominator. Comparing the denominators of (\ref{eq:MGT_p0}) and (\ref{eq:p0b}), and recalling that $\rho$ must satisfy $1 \leq \rho \leq N$, we find that $(2N^2 + \rho W_{s0}^2) > W_{s0}^2$, so $p_0$ of Scenario B will be always smaller than $p_0$ of the Simplified MGT model. Furthermore, empirically, we have noticed that we examined that the real $p_0$ was always lower than $p_0$ based on the Scenario B model with $\rho=1$ in all the cases. If we take this empirical result as always true, it leads to an important conclusion that conservatively setting $\rho=1$, Scenario B always yields a better approximation than the Simplified MGT model.

Note that $W_{s0}$, $q_0$ and $p_0$, as well as $\rho$ are steady state parameters, so their values will be the same for a set of AQM schemes which means that the corresponding  TCP/AQM systems stabilize at the same operating point, while the transient states will distinguish between the dynamics of the different AQM schemes before they converge to steady-state. This is consistent with the results shown in Section~\ref{sec:Numerical}.

\subsection{Different Congestion Levels}
\label{sub:ConLev}

As discussed, our model is based on Scenarios A and B which represent two extreme cases. These two scenarios are definitely not exhaustive as there are many cases where in some TCP sessions \emph{cwnd} $<$ \emph{ssthresh} and in others \emph{cwnd} $\geq$ \emph{ssthresh}, and such cases are not covered by scenarios A and B. To demonstrate that our model is applicable in general, we will need to consider cases that are not included in Scenarios A and B, preferably cases that are very different from Scenarios A and B.

When the congestion level is mild, most TCP sessions are in the congestion avoidance phase, where Scenario B represents the system behavior. When the congestion level increases, the sizes of more and more congestion windows become less than two packets, which means that the \emph{ssthresh} must be equal to 2 (since 2 is the lower bound of \emph{ssthresh}). When most sessions are in the slow start phase, the congestion level is considered to be severe, where Scenario A represents the system behavior. Cases where neither Scenario A nor B can entirely represents the system behavior are characterized by many TCP sessions working in slow start phase and many other sessions working in the congestion avoidance phase.

Having discussed the relationship between Scenarios A and B and the congestion level, it is convenient to introduce a measure of the congestion level, denoted $\overline{w}$, defined by  $\overline{w}=W_{s0}/N$. This measure is the average congestion window of all sessions at the operating point.

When $\overline{w}<1$, most of the TCP sessions have small \emph{cwnd} which is less than 2 packets. In such case, the \emph{cwnd} is even lower than the lower bound of the \emph{ssthresh} which indicate that the majority of TCP sessions work in slow start phase. The link is said to be \emph{severely} congested in this condition. The condition that \emph{cwnd} is greater than the 2 packet lower bound is the necessary condition for a TCP session to work in congestion avoidance phase. The larger $\overline{w}$ is, the more TCP sessions work in congestion avoidance phase. Therefore, we use the condition $\overline{w}>2$ to indicate a region where the link is said to be \emph{mildly} congested where Scenario B accurately represents the system behavior. Having defined the congestion regions of mild and severe, we define the condition $1 \leq \overline{w} \leq 2$ for the region where the link is {\it moderately} congested. Having the two thresholds of $\overline{w}$, i.e, 1 and 2, we now have three non-overlapping and exhaustive regions for the congestion level. Table~\ref{table:congestion level} shows the partition of the three congestion level regions of $\overline{w}$ and Scenarios. As our Scenarios A and B do not cover the moderate congestion region where $1 \leq \overline{w} \leq 2$, we will use Scenarios A and B as the lower and upper bounds to approximate the system behavior.

\begin{table}
\centering
\caption{The Classification of Congestion Level}
\label{table:congestion level}
\begin{tabular}{|c|c|c|c|}
\hline
\diagbox{Average \emph{cwnd} \\ \& Scenario}{Congestion \\ level} & Mild & Moderate & Severe \\
\hline
$\overline{w}$ & $\overline{w}>2$ & $1\leq\overline{w} \leq 2$ & $\overline{w}<1$\\
\hline
Scenario & B & Both A and B & A\\
\hline
\end{tabular}
\end{table}

\subsection{Closeness of the Bounds}
\label{sub:Closeness}
As discussed, Scenarios A and B are applicable to the severe and mild congestion levels, respectively. However, for all the cases in-between these two extreme congestion levels, Scenario A and B models are used as bounds. It is therefore important to discuss their closeness to each other. Here we explain that their closeness can be achieved if we can choose an appropriate $\rho$ parameter for each of the two scenarios and if the system is stable. Then, we show stability for a specific linearized PI AQM system.

Although Scenario A and B use different ways to increase \emph{cwnd}, the queue length and marking probability of the two models converge to the operating point under steady-state conditions if the system is stable. As mentioned, the operating point is defined by $W_{s0}$, $q_0$ and $p_0$. Firstly, the value of $q_0$ depends on the AQM scheme. Hence, the queue length of our model can be stabilized at $q_0$ if the system is stable. Secondly, for a given set of network parameters, the values of $W_{s0}$ of both scenarios are the same according to (\ref{eq:w0}). Let $\rho_A$ and $\rho_B$ be the $\rho$ values for Scenario A and B, respectively.  Then, if the settings satisfy the following equation:
\begin{equation}
\rho_A =\frac{W_{s0}}{N} \rho_B,
\end{equation}
the values of $p_0$ of both scenarios will be the same according to (\ref{eq:p0a}) and (\ref{eq:p0b}). Hence, for a given set of network parameters, Scenario A and B can have the same operating point if a suitable value of $\rho$ is set for each scenario. The models of the two scenarios will converge to the same value under the assumption that the system is stable.

Although it is difficult to prove stability of every TCP/AQM system, we consider a particular system based on PI AQM under the settings used in Section ~\ref{sec:Numerical} and in Appendix~\ref{sec:app1} we provide stability analysis and numerical verification for this TCP/AQM system for the case where the system is linearized at the operating point. Routh-Hurwitz stability criterion is then used to determine whether this TCP/AQM system is stable.

\subsection{The Parameter $\rho$}
\label{sub:rho}

We have already discussed the importance of setting $\rho$ to achieve accuracy and closeness of the bounds. As discussed, we do not have an analytical method to obtain it. Nevertheless, we know that it is bounded within $1 \leq \rho \leq N$ and that simply and conservatively setting $\rho = 1$ gives improved results over the Simplified MGT model, which will be demonstrated through a large sample of cases over a wide range of parameters in the next section. To appreciate the difficulty in obtaining $\rho$, notice that $\rho$ could have been obtained by (\ref{eq:w0}), (\ref{eq:RTT_op}), (\ref{eq:p0a}), and (\ref{eq:p0b}) if $p_0$ is available analytically. However, obtaining $p_0$ analytically is a well known open problem. By the same equations, if $\rho$ is analytically available then $p_0$ can be obtained. This explain that obtaining $\rho$ is equivalent to the known difficult problem of finding $p_0$.

One way is to obtain $p_0$ (and $\rho$) is by simulation. Such values will be used in many simulations in the next section to demonstrate how accurate our model could be if we were able to obtain $\rho$ analytically. As discussed, the results for the cases of $\rho=1$ will also be discussed and compared with. As mentioned, $\rho$ is a steady-state parameter, therefore if the parameter $\rho$ is correct for one AQM scheme, it can be reused for the other AQM schemes who has the same operating point. This property will be illustrated in the next section.

\section{NUMERICAL RESULTS}
\label{sec:Numerical}
In this section, we show the improvement of the proposed model compared to the Simplified MGT model for two settings of the parameter $\rho$: one is the conservative case of $\rho=1$ and the other is the case where $\rho$ is obtained based on the NS2 simulation results.

We first introduce the network topology used for the NS2 simulation, followed by information on the setting of the default parameters of three AQM schemes: PI, REM and RaQ. Then, we conduct a series of simulations to evaluate the new model by comparing the analytical results with simulation results. All the time-dependent analytical results are based on the discrete-time model of Scenarios A and B and a discrete time version of the Simplified MGT model. In all the figures in this section that show comparative simulation results, for clarity, we use the label Simplified MGT to represent the results obtained for the Simplified MGT model, we use the labels Scenario A, Scenario B, with $\rho = 1$ to represent the analytical results obtained for Scenarios A and B setting $\rho = 1$, and equivalently we use the labels Scenario A, Scenario B together with the specific $\rho$  value obtained from the NS2 simulations to represent an ideal analytical results for Scenarios A and B indicating the limit of possible improvement based on our model. We observe in all the numerical results that all Scenario A and B curves (including the cases of $\rho=1$) converge to the target queue length. This is consistent with our discussions in Subsections \ref{sub:Closeness}.

\subsection{Experiment Topology and Default Parameters}
\label{sub:topology}

\begin{figure}[!h]
\centering
\includegraphics[width=\linewidth]{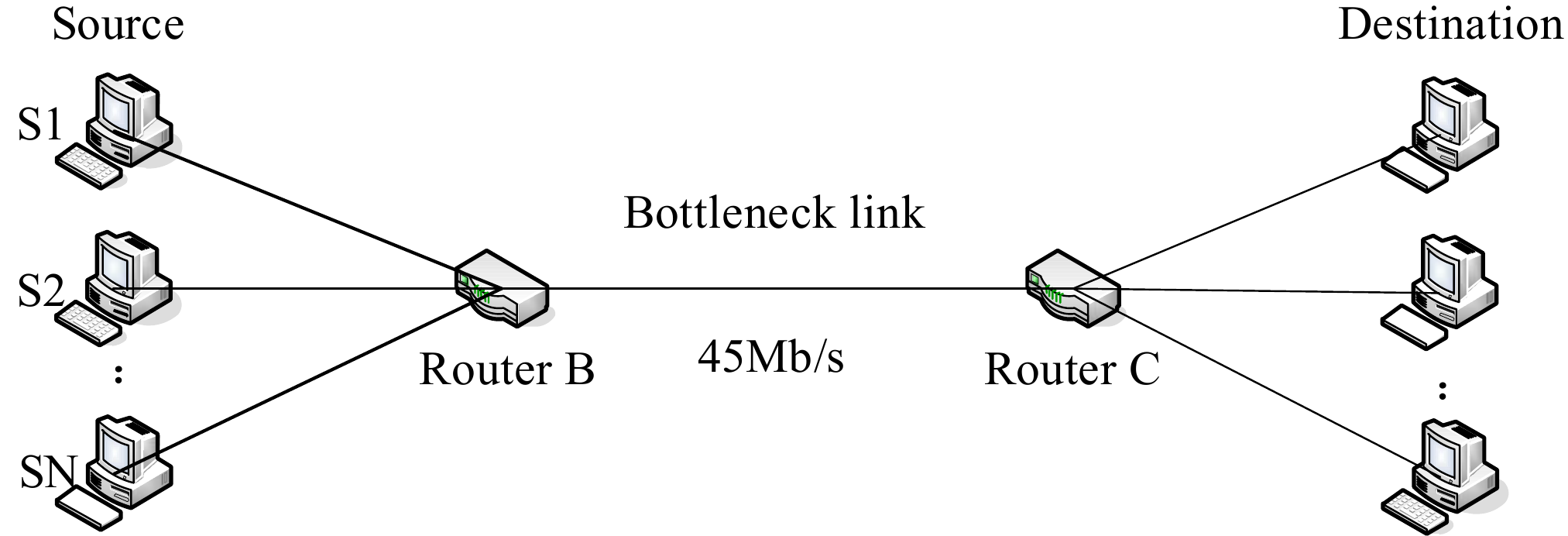}
\caption{The topology of simulation network.}
\label{fig:topology}
\end{figure}

In our simulations, we use a dumb-bell network topology shown in Fig.~\ref{fig:topology}. The link between router B and router C is the bottleneck link. Unless mentioned otherwise, the following parameters are set as default: the mean packet size is 1000 bytes, the propagation time is 100 ms, the bottleneck link capacity is 45 Mb/s, the buffer size is 1125 packets, the target queue length is 500 packets, and the number of TCP sessions is 500. TCP/Reno is adopted as the TCP plant and ECN is set ON unless otherwise stated.

The default parameters of the AQM schemes are set as follows: For PI, $a = 0.00001822$ and $b = 0.00001816$. For REM, $\gamma = 0.001$ and $\Phi=1.001$. For RaQ, $q_{kp} = 0.0077$, $q_{ki} = 0.0005$, and $r_{kp} = 0.0095$. The above PI \cite{Hollot2001PI} and REM \cite{Athuraliya:2001:REM} parameters are the NS2 defaults and the RaQ parameters are as in \cite{ Sun2007RaQ}.

The default sampling period $\Delta t$ used for the analytical results for Scenarios A and B and the Simplified MGT model is 0.0005 s, and the default sampling period $T$ used for the three AQM schemes PI, REM and RaQ is 0.005 s. Note that $\Delta t$ should be smaller than or equal to $T$. Further discussion of these two sampling periods will be given in Subsection \ref{subsec:step time} below.

\subsection{Performance Under Different Congestion Levels}
\label{sub:Perfo_ConLev}
As discussed, our analytical model that are based on Scenario A and B should be tested for a wide range of congestion levels. To this end, we consider five cases, three of which represent the three congestion levels: mild, severe and moderate, and the other two represent borderline cases \emph{mild/moderate} and \emph{moderate/severe} that approximately correspond to the cases $\overline{w}=2$ and $\overline{w}=1$, respectively. These five cases are described in Table~\ref{table:simulation cases}. The number of TCP sessions $N$ is adjusted to achieve different congestion levels. In particular, we use the $N$ values of 200, 500, 800, 1100 and 2000. For example, in the first case, of $N = 200$, $\overline{w} = CR_0/N=5625*0.1889/200 = 5.3128$, which is more than 2 packets and therefore, this case is within the mild congestion level region.

\begin{table}[!h]
\centering
\caption{Congestion Levels and the Corresponding $N$}
\label{table:simulation cases}
\begin{tabular}{|c|c|c|}
\hline
$N$ & $\overline{w}$ & Congestion level \\
\hline
200 & 5.3128 & Mild\\
\hline
500 & 2.1251 & Mild/Moderate\\
\hline
800 & 1.3282 & Moderate\\
\hline
1100 & 0.9660 & Moderate/Severe\\
\hline
2000 & 0.5313 & Severe\\
\hline
\end{tabular}
\end{table}

\subsubsection{Mild Congestion}

\begin{figure}[!h]
\centering
\includegraphics[width=0.5\textwidth]{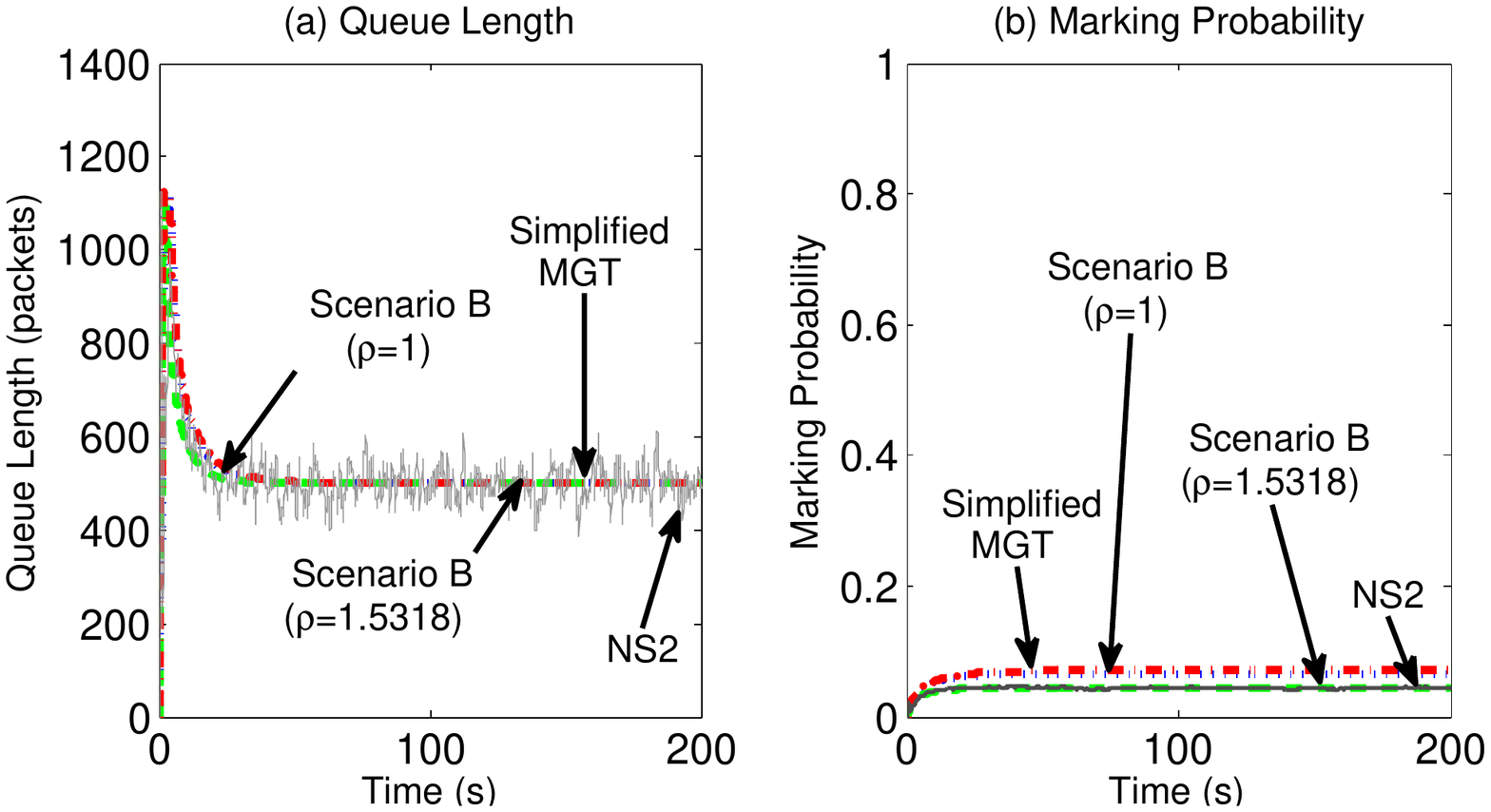}
\caption{Queue length and marking probability versus time for PI AQM with $N = 200$.}
\label{fig:PI200}
\end{figure}

\begin{figure}[!h]
\centering
\includegraphics[width=0.5\textwidth]{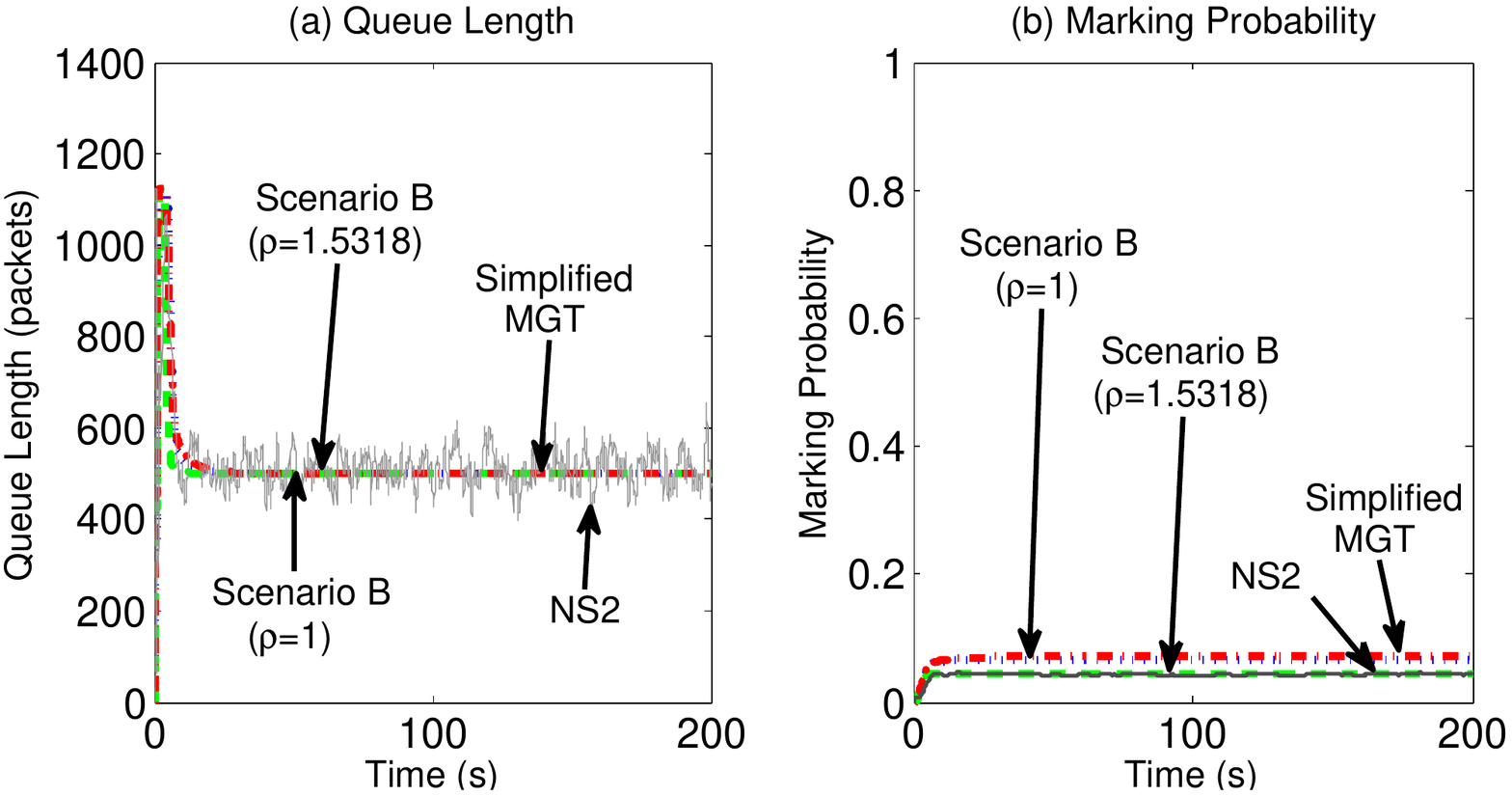}
\caption{Queue length and marking probability versus time for REM AQM with $N = 200$.}
\label{fig:REM200}
\end{figure}

\begin{figure}[!h]
\centering
\includegraphics[width=0.5\textwidth]{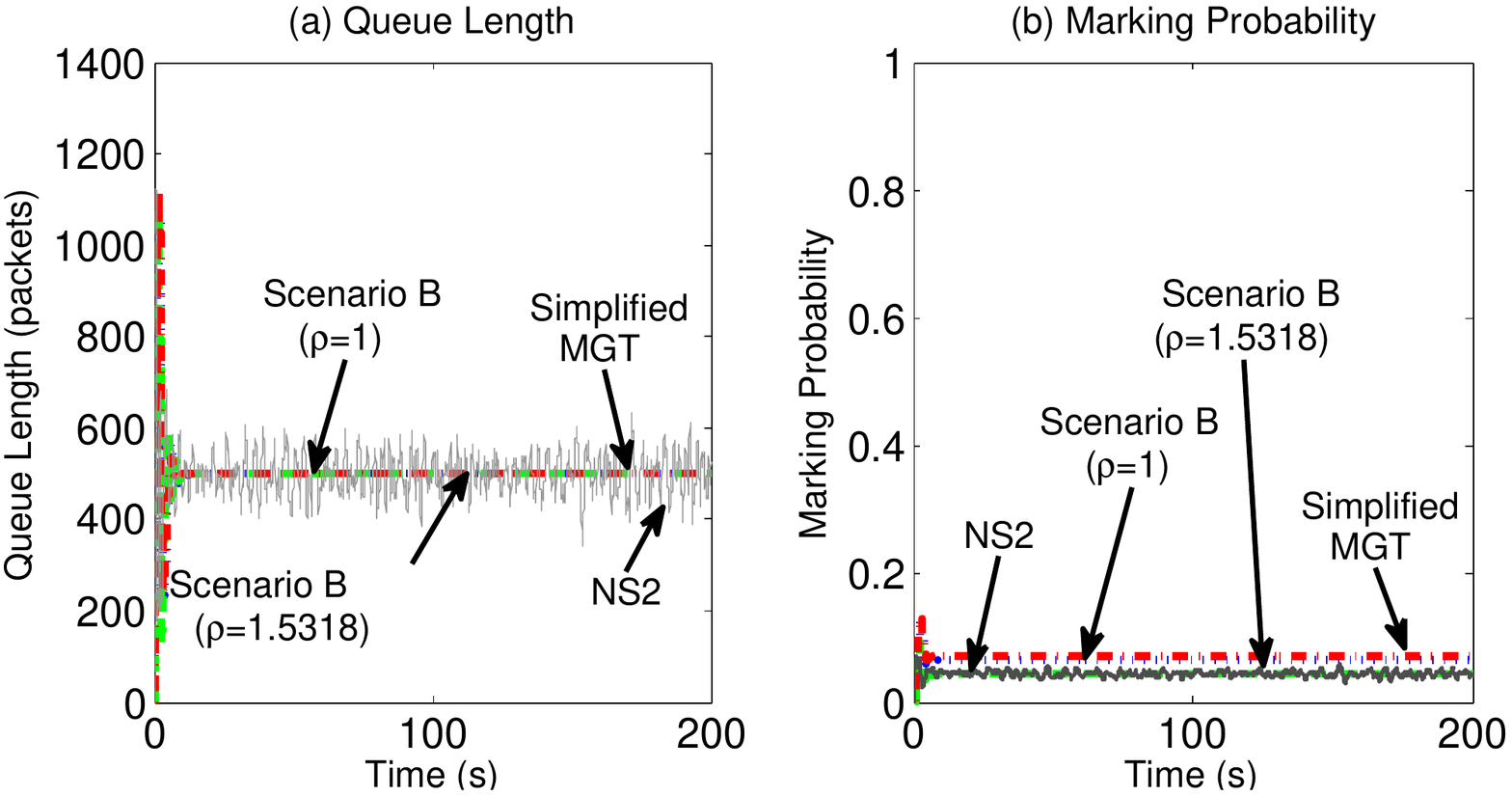}
\caption{Queue length and marking probability versus time for RaQ AQM with $N = 200$.}
\label{fig:RaQ200}
\end{figure}

In this case, we consider PI as the AQM scheme and $N=200$. As this case falls within the mild congestion region, it is well modeled by Scenario B. Based on NS2 simulations, we obtain  $p_0 = 0.0442$, which gives by (\ref{eq:SceB}), $\rho = 1.5318$. By (\ref{eq:MGT_p0}) and (\ref{eq:p0b}), the $p_0$ values of the Simplified MGT model and the Scenario B model (with $\rho=1$) are given in Table~\ref{table:p0_Mild}. We reuse the value $\rho=1.5318$ in the other two AQM schemes REM and RaQ. Results for all three AQM schemes are presented in Fig.~\ref{fig:PI200}, Fig.~\ref{fig:REM200} and Fig.~\ref{fig:RaQ200}, for PI, REM and RaQ, respectively. We observe that the results of all analytical models for all three AQM schemes are fairly accurate in both in steady state and during the transient periods before steady state. For example, all analytical results for all three AQM schemes exhibit convergence to the queue length target of 500 packets.  In addition, we notice for all three AQM schemes that the marking probability dynamics based on Scenario B with $\rho=1$ is somewhat more accurate than the results obtained by the  Simplified MGT model which is consistent with the steady state results shown in Table~\ref{table:p0_Mild}.

\begin{table}[!h]
\centering
\caption{Values of $p_0$ for $N = 200$.}
\label{table:p0_Mild}
\begin{tabular}{|c|c|c|c|}
\hline
$N$ & \tabincell{c}{Simplified \\ MGT} & \tabincell{c}{Scenario B \\ ($\rho=1$)}& NS2 \\
\hline
200 & 0.0708 & 0.0662 & 0.0442\\
\hline
\end{tabular}
\end{table}

\subsubsection{Severe Congestion}

\begin{figure}[!h]
\centering
\includegraphics[width=0.5\textwidth]{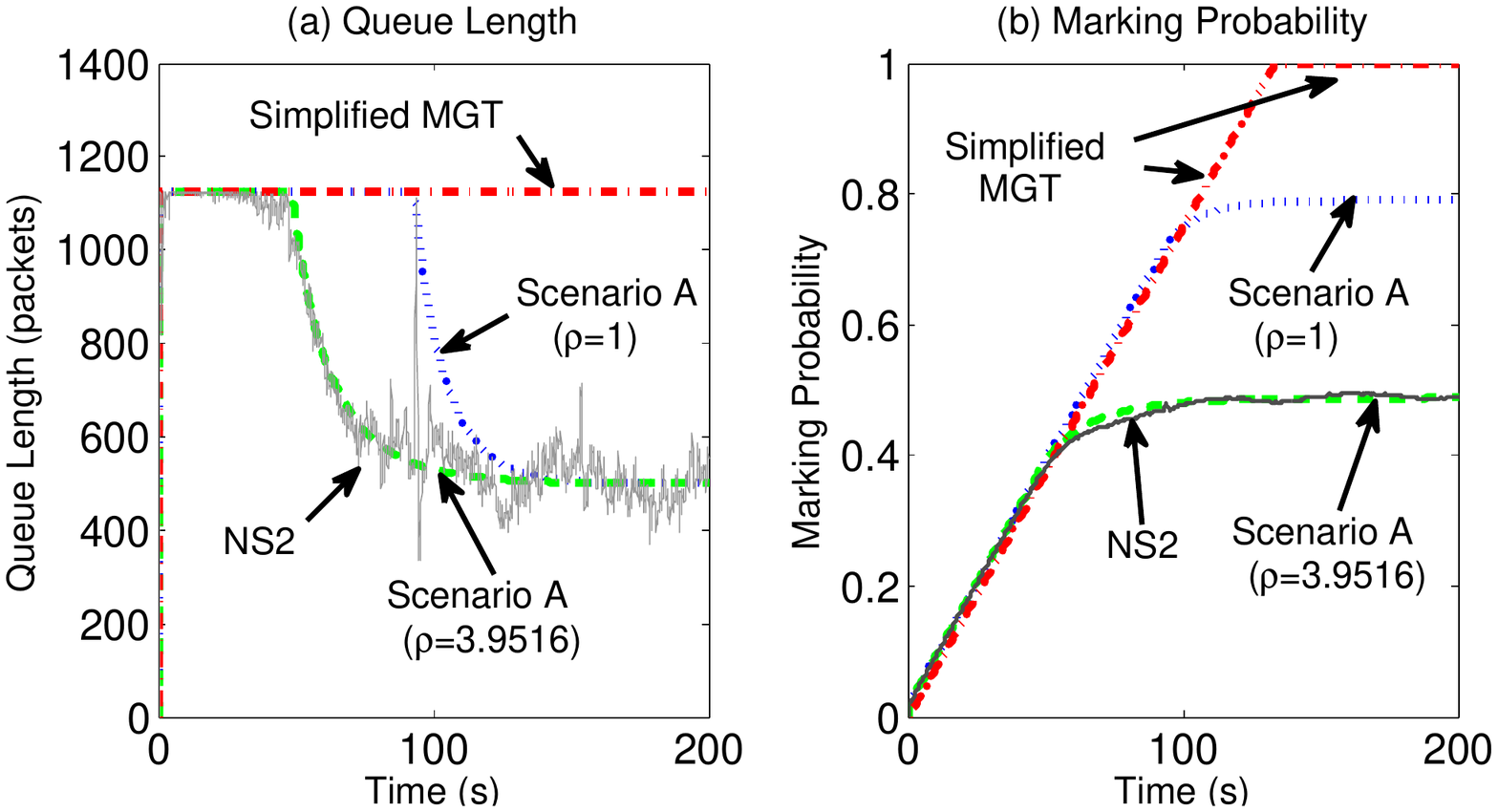}
\caption{Queue length and marking probability versus time for PI AQM with $N = 2000$.}
\label{fig:PI2000}
\end{figure}

\begin{figure}[!h]
\centering
\includegraphics[width=0.5\textwidth]{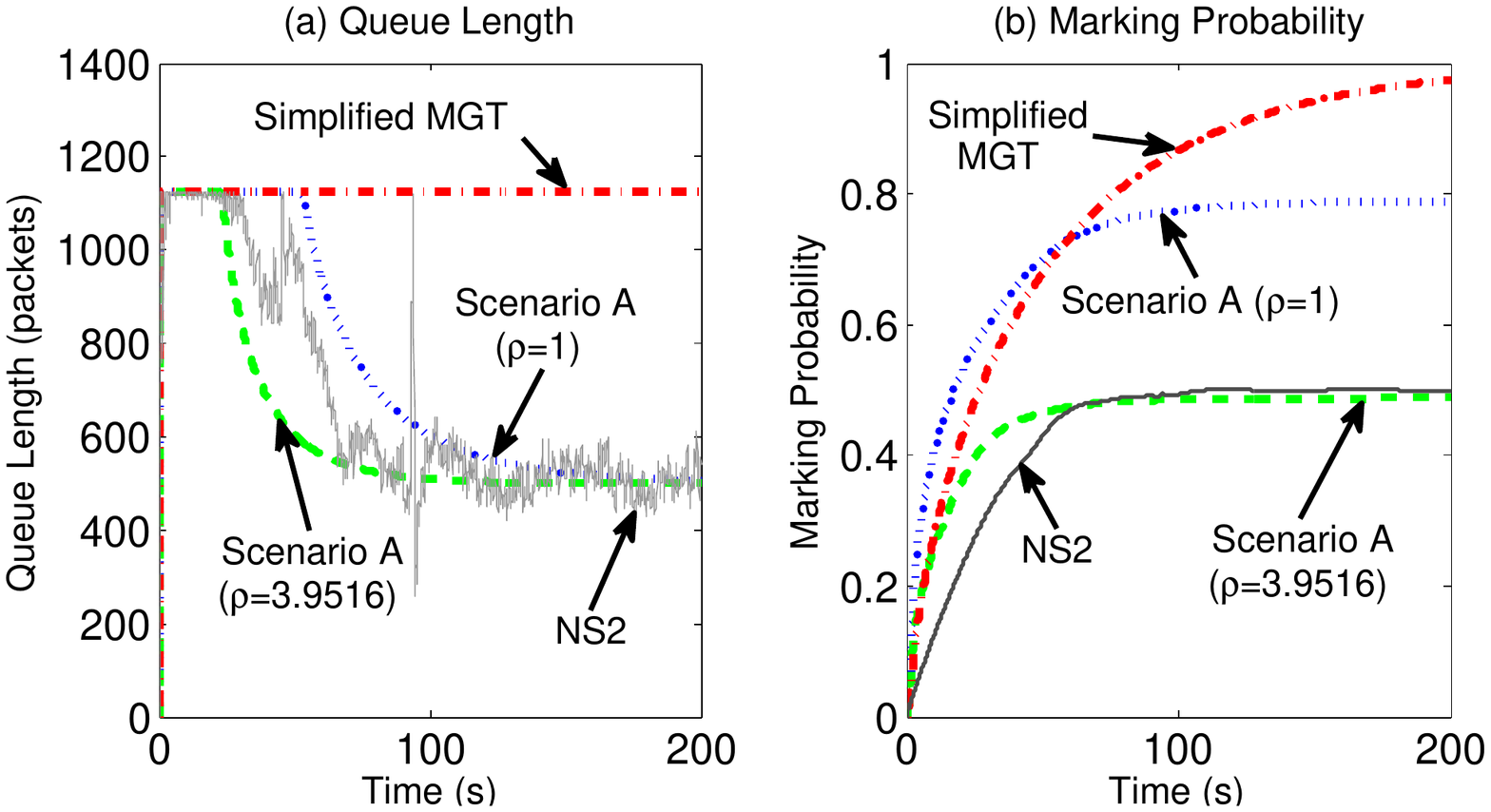}
\caption{Queue length and marking probability versus time for REM AQM with $N = 2000$.}
\label{fig:REM2000}
\end{figure}

\begin{figure}[!h]
\centering
\includegraphics[width=0.5\textwidth]{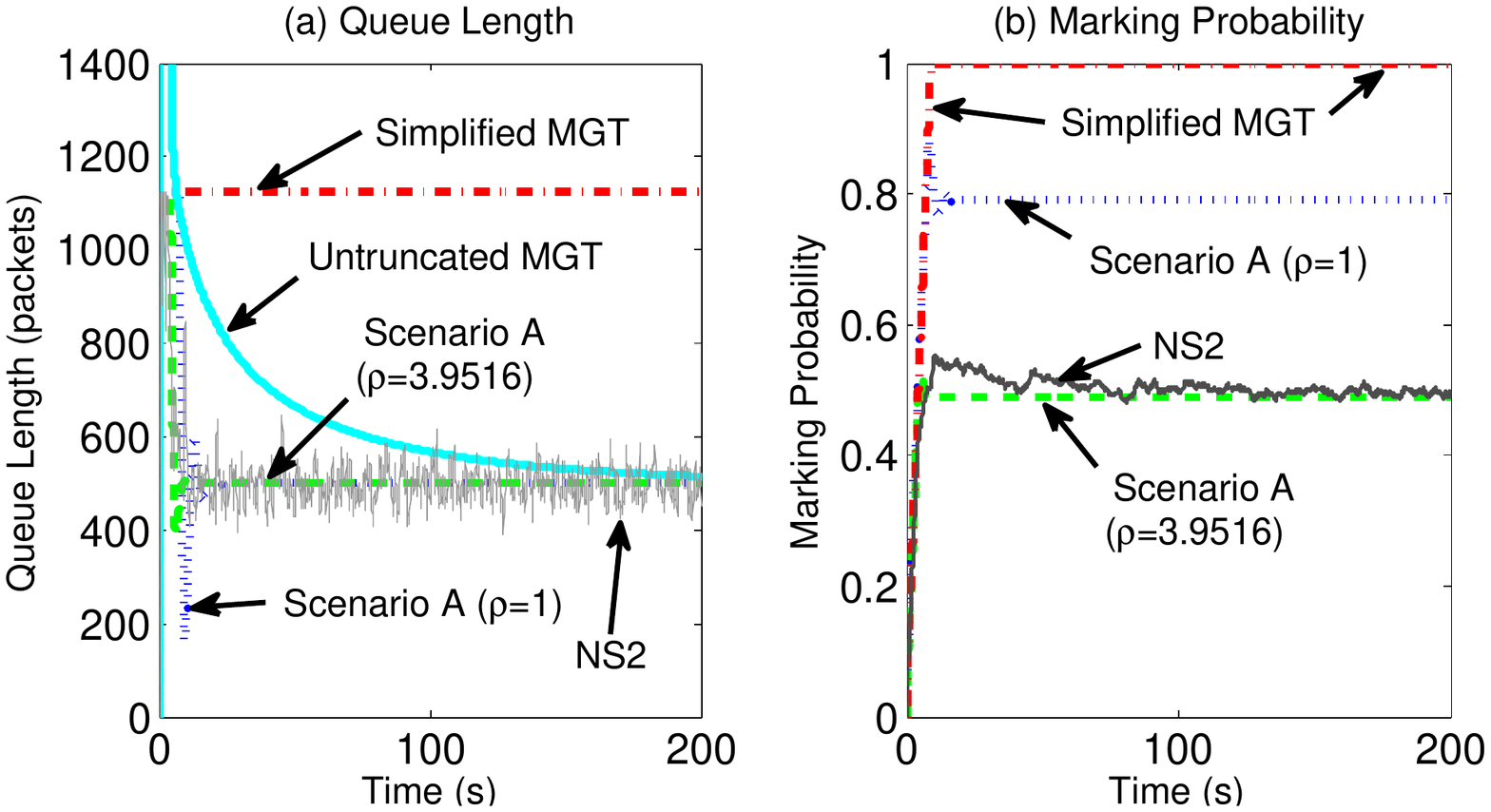}
\caption{Queue length and marking probability versus time for RaQ AQM with $N = 2000$.}
\label{fig:RaQ2000}
\end{figure}

In this case, we again first consider PI as the AQM scheme and we set $N=2000$. This case falls within the severe congestion region, so it is modeled by Scenario A. Based on NS2 simulations, we obtain $p_0 = 0.4879$, which gives by (\ref{eq:SceA}), $\rho = 3.9516$. By (\ref{eq:MGT_p0}) and (\ref{eq:p0a}), the $p_0$ values of the Simplified MGT model and Scenario A (with $\rho=1$) are given in Table~\ref{table:p0_Severe}. We notice that the $p_0=0.7901$ predicted by Scenario A (with $\rho=1$) improves the $p_0=7.0827$ of the Simplified MGT model, while still it is not close to the NS2 $p_0 = 0.4879$. We reuse the value $\rho=3.9516$ in the other two AQM schemes REM and RaQ. Results for all three AQM schemes are presented in Fig.~\ref{fig:PI2000}, Fig.~\ref{fig:REM2000} and Fig.~\ref{fig:RaQ2000}, for PI, REM and RaQ, respectively.

We observe that the results of Scenario A (with $\rho=1$) for all three AQM schemes improve the accuracy both in steady state and during the transient periods. We observe in Fig.~\ref{fig:PI2000} (a), that although the curve of Scenario A ($\rho=1$) does not very accurately predict the transient period, it converges to the target queue length of 500 packets, despite the fact that the $p_0$ is inaccurate.  By comparison, the Simplified MGT model predicts that the queue length saturates the buffer.  As for the marking probability dynamics, the curve of Scenario A (with $\rho=1$) converges to the value 0.7901 after 100 second, while the Simplified MGT model converges to 1 only because it is truncated to 1. In the other two figures associated with REM and RaQ AQM, Scenario A (with $\rho=1$) again converges to the target queue length and provides more accurate marking probability than the Simplified MGT model. The results also demonstrate that Scenario A (with $\rho=3.9516$) provides curves closest to the curves of the NS2 simulations by comparison to the Simplified MGT model and Scenario A (with $\rho=1$) for both queue length dynamics and marking probability.

\begin{figure}[!h]
\centering
\includegraphics[width=0.5\textwidth]{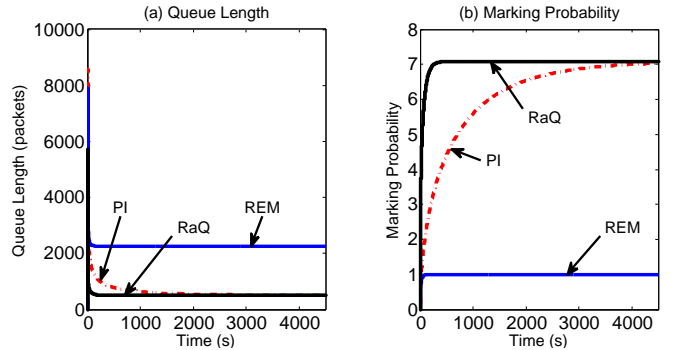}
\caption{Queue length and marking probability of untruncated MGT model for PI, REM and RaQ schemes with $N=2000$.}
\label{fig:untruncated}
\end{figure}

We also presented the result of MGT model without truncation (for both marking probability and queue length) in Fig.~\ref{fig:untruncated}. It is shown that the queue length for RaQ or PI can converge to the target queue length of 500 packets, while the marking probability is stabilized to more than 7. However, the marking probability of REM does not exceed 1 because its calculation is based on the function $1- \Phi^{-pl}$, but the consequence is that the queue length cannot converge to the target queue length. We also present the queue length dynamics of the Simplified MGT model without truncation in Fig.~\ref{fig:RaQ2000}. We observe that although some improvement in the modeling of queue length dynamic is achieved without truncation, relative to the truncated counterpart, the accuracy under severe congestion is not satisfactory. Note that the convergence of the untruncated Simplified MGT for PI takes 2000 seconds for the queue size and 4000 for the marking probability, so the results are not included in Fig.~\ref{fig:PI2000}. We also exclude the untruncated MGT results from Fig.~\ref{fig:REM2000} because they do not improved on their truncated counterparts.

\begin{table}[!h]
\centering
\caption{Values of $p_0$ for $N=2000$}
\label{table:p0_Severe}
\begin{tabular}{|c|c|c|c|}
\hline
$N$ & \tabincell{c}{Simplified \\ MGT} & \tabincell{c}{Scenario A \\ ($\rho=1$)}& NS2 \\
\hline
2000 & 7.0827 & 0.7901 & 0.4879\\
\hline
\end{tabular}
\end{table}

\subsubsection{Mild/Moderate, Moderate and Moderate/Severe Congestion}

\begin{figure}[!h]
\centering
\includegraphics[width=0.5\textwidth]{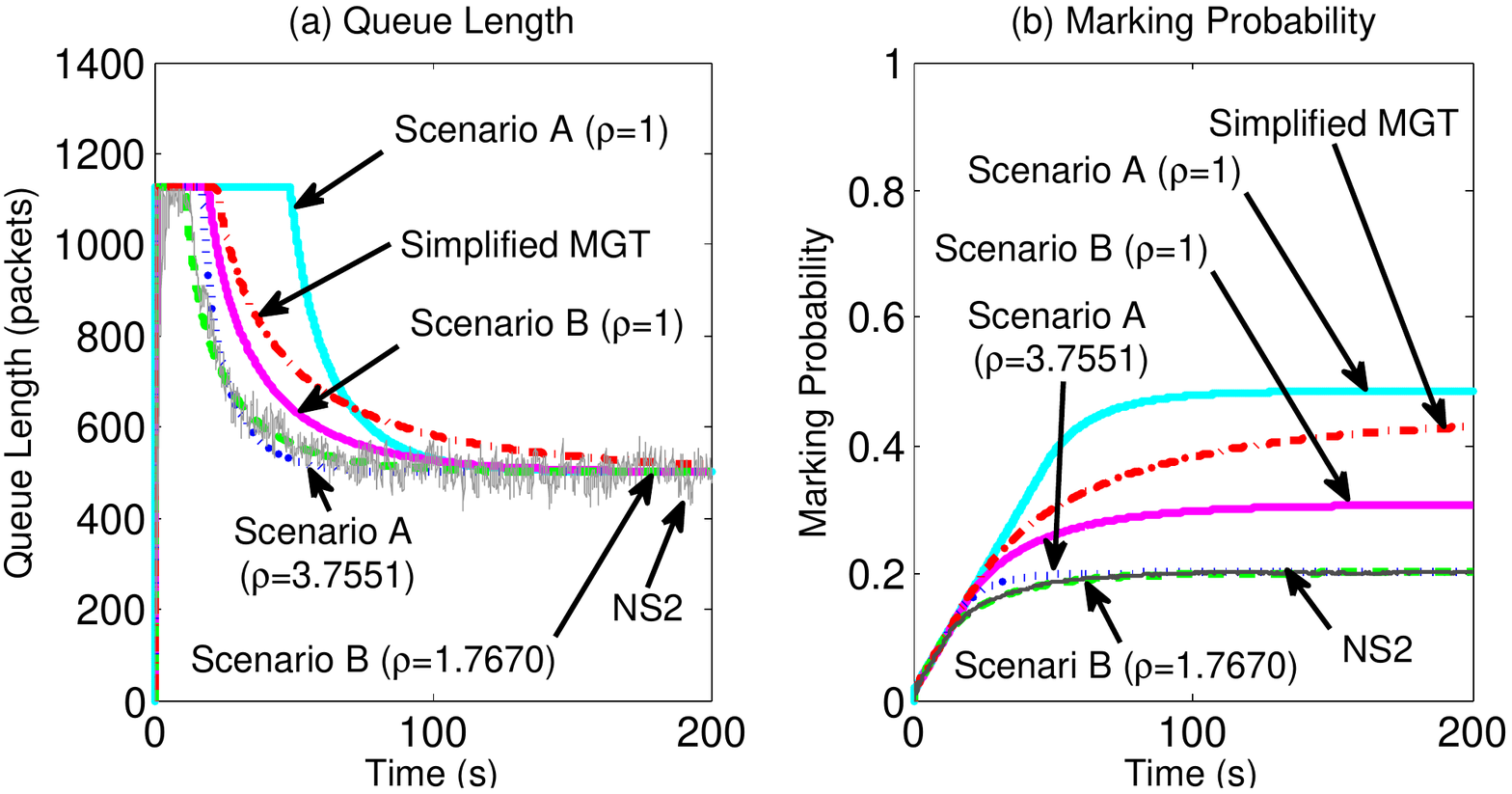}
\caption{Queue length and marking probability versus time for PI AQM with $N = 500$.}
\label{fig:PI500}
\end{figure}

\begin{figure}[!h]
\centering
\includegraphics[width=0.5\textwidth]{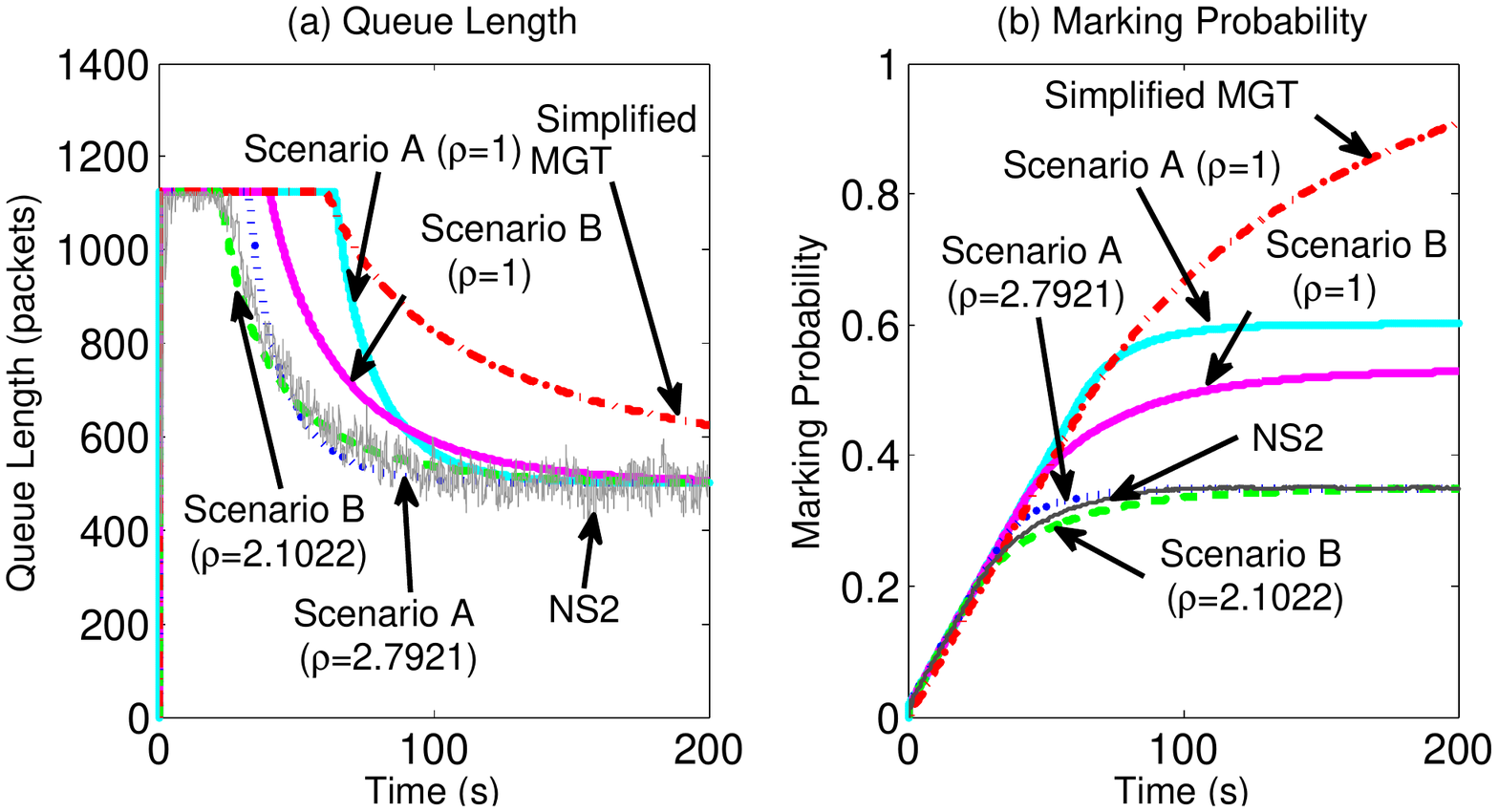}
\caption{Queue length and marking probability versus time for PI AQM with $N = 800$.}
\label{fig:PI800}
\end{figure}

\begin{figure}[!h]
\centering
\includegraphics[width=0.5\textwidth]{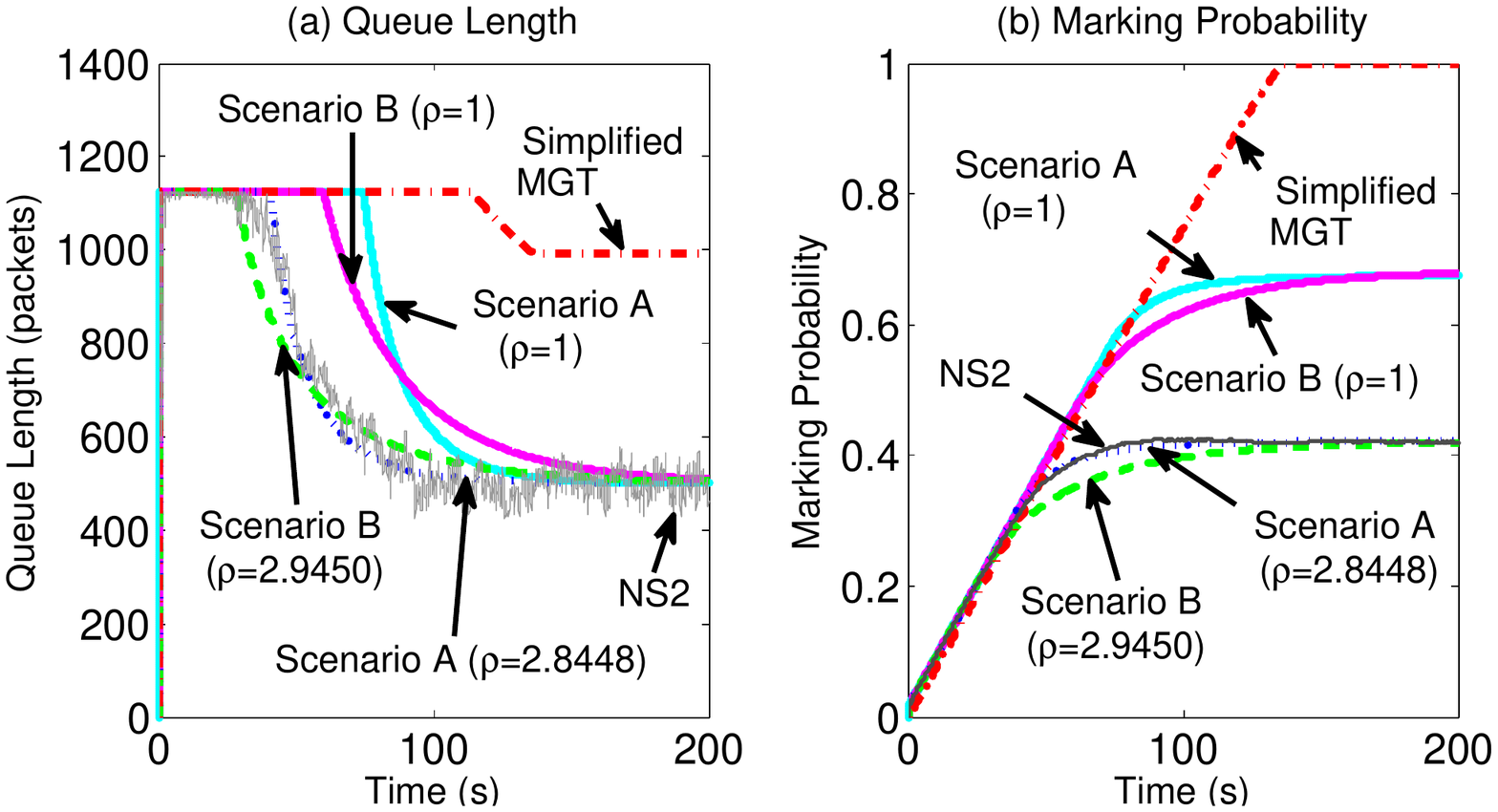}
\caption{Queue length and marking probability versus time for PI AQM with $N = 1100$.}
\label{fig:PI1100}
\end{figure}

We now consider the congestion levels associate with the cases $N=500, 800, 1100$. For each case, we obtain $p_0$ from the NS2 simulations of PI AQM, and derive the $\rho$ values from (\ref{eq:p0a}) and (\ref{eq:p0b}) provided in Table~\ref{table:ro_moderate}. We also obtain $p_0$ of the Simplified MGT model and Scenarios A and B for $\rho=1$ by (\ref{eq:MGT_p0}), (\ref{eq:p0a}) and (\ref{eq:p0b}), respectively. The values are presented in Table~\ref{table:p0_moderate}.
Fig.~\ref{fig:PI500}, Fig.~\ref{fig:PI800} and Fig.~\ref{fig:PI1100} show the numerical results of the NS2 simulation and of the analytical models. Comparing to the results of NS2 simulations, it can be found that the convergence time of queue length and marking probability at the steady state increase when $N$ increases. For the cases that $\rho$ are set to the values based on NS2 simulation (Table~\ref{table:ro_moderate}), the results show that, as expected, Scenario B performs better than Scenario A near the bound of Mild/Moderate, while Scenario A performs better near the bound of Moderate/Severe. In the moderate congestion range, the curves of NS2 simulations are between the Scenario B and Scenario A curves. Therefore, these results confirm that the milder the congestion level is, the better Scenario B model performs, and vice versa. This conclusion also applies to the cases where $\rho=1$.  In addition, Scenario A and B curves are approaching each other in the cases where the $\rho$ values are set based on the NS2 simulations. This is consistent with the discussion in Subsection~\ref{sub:Closeness}. Also Scenario A and B curves with $\rho=1$ are reasonable close to each other for queue dynamic results of the three AQM schemes.

\begin{table}[!h]
\centering
\caption{ Values of $p_0$ for $N = 500, 800, 1100$}
\label{table:p0_moderate}
\begin{tabular}{|c|c|c|c|c|}
\hline
$N$ & \tabincell{c}{Simplified \\ MGT} & \tabincell{c}{Scenario B \\ ($\rho=1$)}& \tabincell{c}{Scenario A \\ ($\rho=1$)} & NS2 \\
\hline
500 & 0.4429 & 0.3069 & 0.4848 & 0.2004\\
\hline
800 & 1.1337 & 0.5313 & 0.6009 & 0.3504\\
\hline
1100 & 2.1434 & 0.6819 & 0.6743 & 0.4212\\
\hline
\end{tabular}
\end{table}

\begin{table}[!h]
\centering
\caption{ Parameter $\rho$ Of New Model For $N = 500, 800, 1100$ Based On NS2 Simulations}
\label{table:ro_moderate}
\begin{tabular}{|c|c|c|}
\hline
$N$ &  $\rho$ for Scenario B & $\rho$ for Scenario A \\
\hline
500 & 1.7670 & 3.7551\\
\hline
800 & 2.1022 & 2.7921\\
\hline
1100 & 2.9450 & 2.8448\\
\hline
\end{tabular}
\end{table}

\subsection{Turning OFF ECN}
In the case where ECN is turned off, by NS2 simulations, we obtain $p_0 = 0.1416$. The congestion level index is $\overline{w} = CR_0/(N-Np_0) = 2.4756$. By Table~\ref{table:congestion level}, such congestion level falls into the mildly congested region, so the model of Scenario B is adopted. Then from (\ref{eq:w0}) and (\ref{eq:p0a}), we obtain
\begin{equation}
\rho = \frac{(1-p_0)^3 * 2N^2}{C^2 R_0^2 p_0} = 1.9789.
\end{equation}

Since the Simplified MGT model does not specifically consider ECN, the previous case of $N=500$, which yields $p_0 (\rm MGT) = 0.4429$ is used here. While the $p_0$ of Scenario B model ($\rho=1$) when turning off ECN is obtained by (\ref{eq:w0}) and (\ref{eq:p0b}):

\begin{equation}
\label{eq:p_0b_ECN_OFF}
p_0 (Scenario B, ECN\_OFF)= \dfrac{2N^2}{2N^2 + \dfrac{R_{0}^{2}C^{2}}{(1-p_0)^2}} .
\end{equation}

\begin{figure}[!h]
\centering
\includegraphics[width=0.5\textwidth]{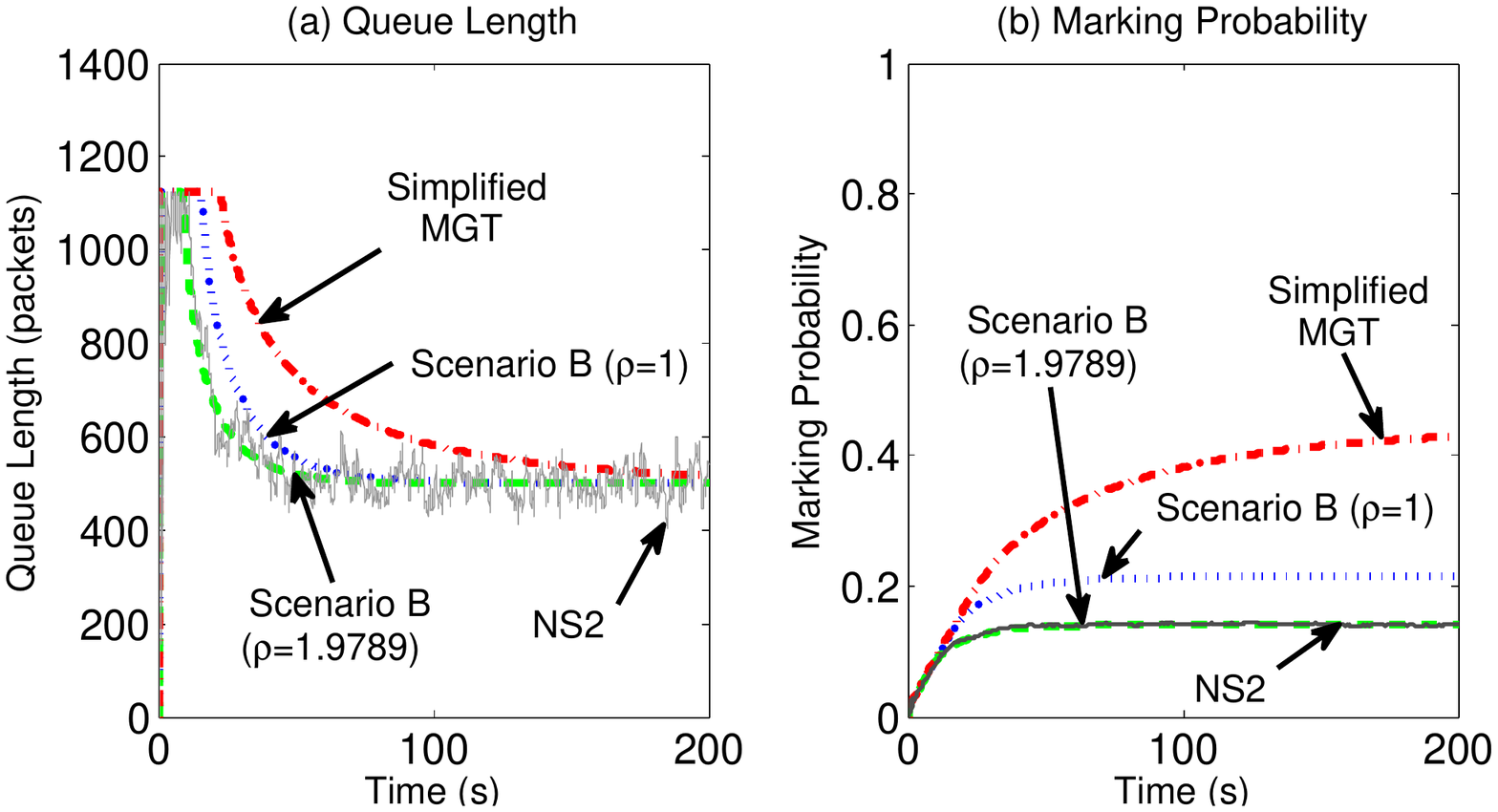}
\caption{Queue length and marking probability versus time for PI AQM with $N = 500$ when ECN is OFF.}
\label{fig:PI500drop}
\end{figure}

\begin{figure}[!h]
\centering
\includegraphics[width=0.5\textwidth]{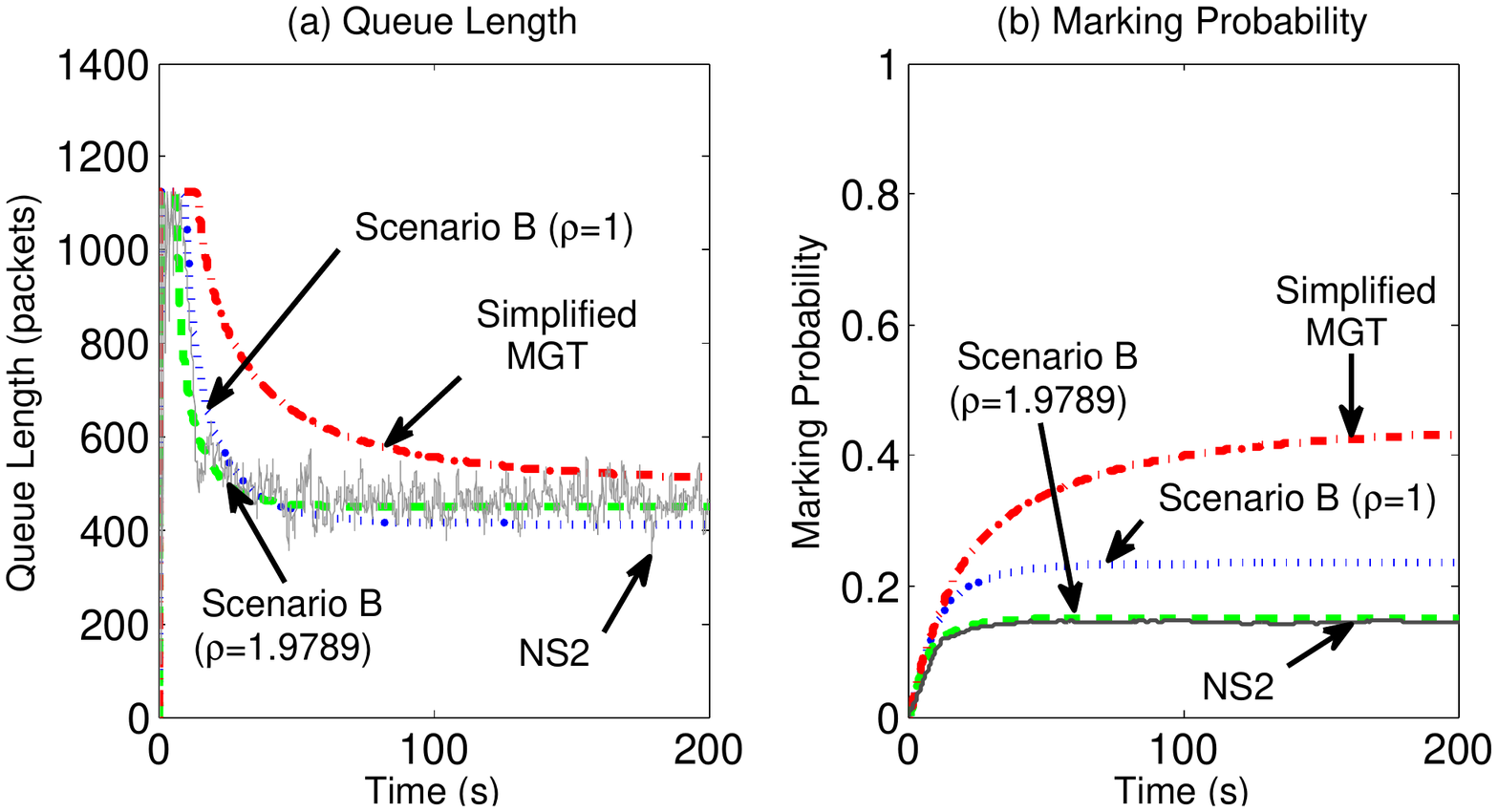}
\caption{Queue length and marking probability versus time for REM AQM with $N = 500$ when ECN is OFF.}
\label{fig:REM500drop}
\end{figure}

\begin{figure}[!h]
\centering
\includegraphics[width=0.5\textwidth]{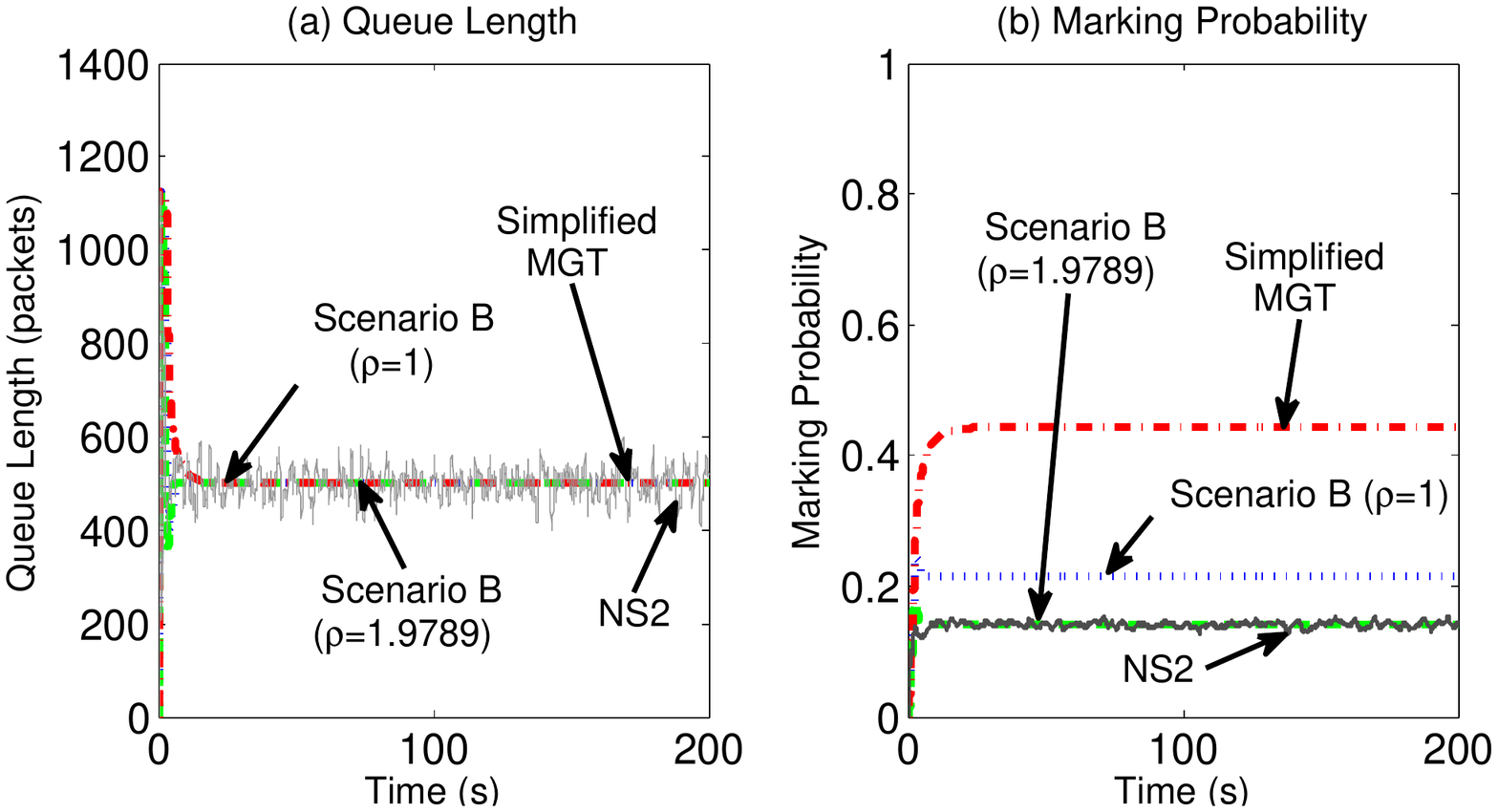}
\caption{Queue length and marking probability versus time for RaQ AQM with $N = 500$ when ECN is OFF.}
\label{fig:DEM500drop}
\end{figure}

Isolating $p_0$ and solving this cubic equation, we obtain $p_0 (Scenario B, \rho=1)= 0.2146$. The simulation and analytical results are shown in Fig.~\ref{fig:PI500drop}, Fig.~\ref{fig:REM500drop} and Fig.~\ref{fig:DEM500drop}.

The results exhibit that the curves of Scenario B (with $\rho=1$) are closer to the curves of NS2 simulations in both Fig.~\ref{fig:PI500drop} and Fig.~\ref{fig:REM500drop} compared to the Simplified MGT model¡¯s results. The marking probabilities are consistent with the theoretical value calculated by  (\ref{eq:p_0b_ECN_OFF}).

Note that the target queue length is set as 500 packets in NS2 simulations. In Fig.~\ref{fig:REM500drop}, where REM AQM is adopted, the stable queue length of the NS2 simulation is around 450 packets, which is lower than the target queue length (500 packets). In this case, Scenario B ($\rho=1.9789$) can still captures this discrepancy, while the queue length of the Simplified MGT model converges to 500 packets. This fact demonstrates that the correct parameter $\rho$ is reusable under the stability assumption, regardless of which AQM scheme is used. For the other two AQM schemes, Scenario B model ($\rho=1.9789$) provide accurate match for both the queue length and dropping probability dynamic.

\subsection{Varying Number of TCP Sessions}
This set of simulations tests the ability of tracking a link where the number of TCP sessions is varied. Specifically, 300 TCP sessions start at the time 0, 200 additional sessions join at 65 s, and 200 TCP sessions stop at 130 s. According to (\ref{eq:w0}), (\ref{eq:RTT_op}) and (\ref{eq:p0b}) and based on NS2 simulation, we have the values of $\rho = 1.6575$ and $\rho = 1.7670$ when $N = 300$ and $N = 500$, respectively. By (\ref{eq:MGT_p0}) and (\ref{eq:p0b}), the theoretical $p_0$ of Scenario B ($\rho=1$) and the Simplified MGT model for $N=300$ and $N=500$ are given in Table~\ref{table:p0_varying}.

\begin{table}[!h]
\centering
\caption{ Values of $p_0$ for $N$ Changes}
\label{table:p0_varying}
\begin{tabular}{|c|c|c|c|}
\hline
$N$ & \tabincell{c}{Simplified \\ MGT} & \tabincell{c}{Scenario B \\ ($\rho=1$)} & NS2 \\
\hline
300 & 0.1594 & 0.1375 & 0.0877 \\
\hline
500 & 0.4429 & 0.3069 & 0.2004 \\
\hline
\end{tabular}
\end{table}

\begin{figure}[!h]
\centering
\includegraphics[width=0.5\textwidth]{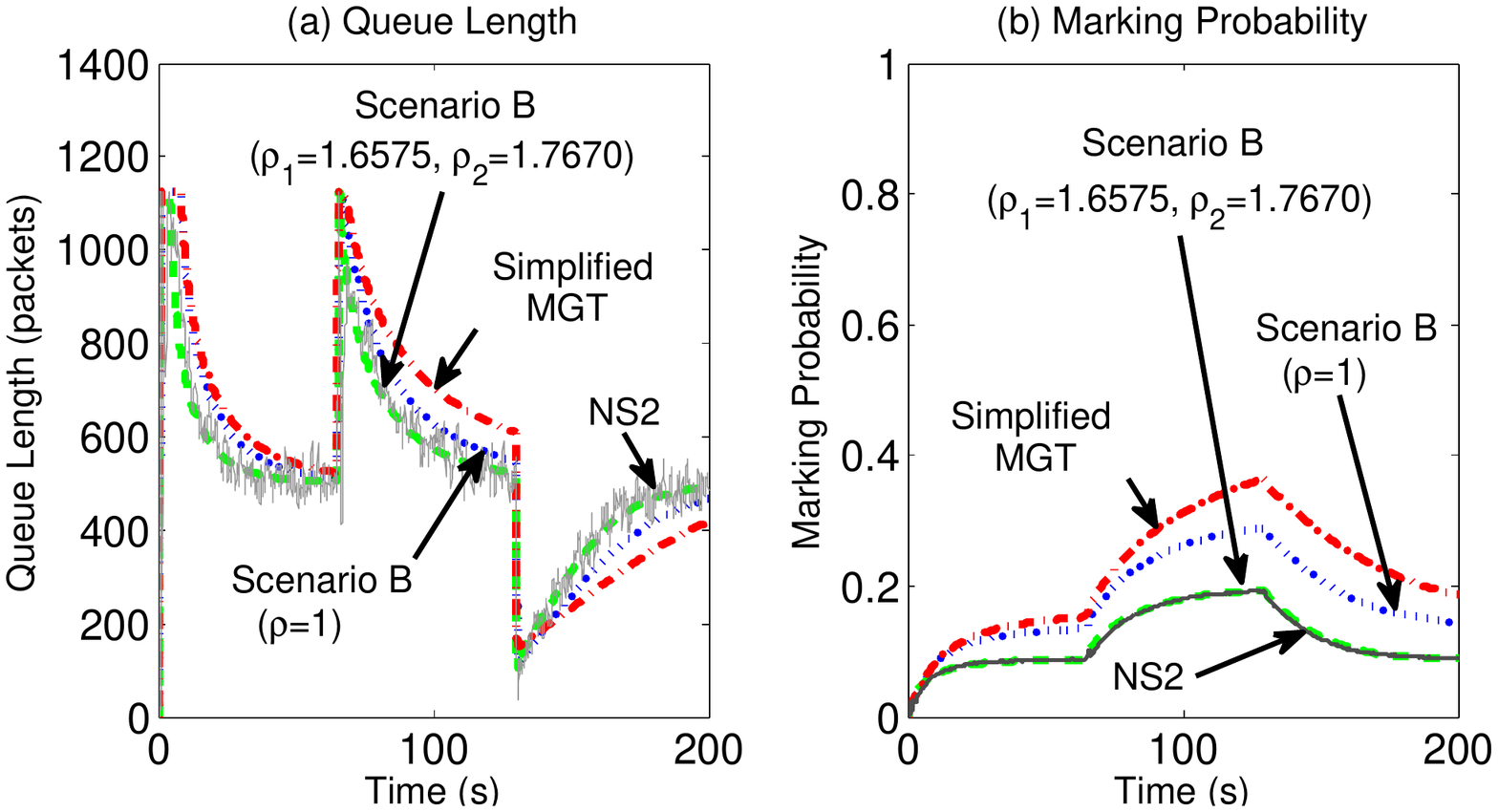}
\caption{Queue length and marking probability versus time for PI AQM when $N$ varies.}
\label{fig:PI300200}
\end{figure}

\begin{figure}[!h]
\centering
\includegraphics[width=0.5\textwidth]{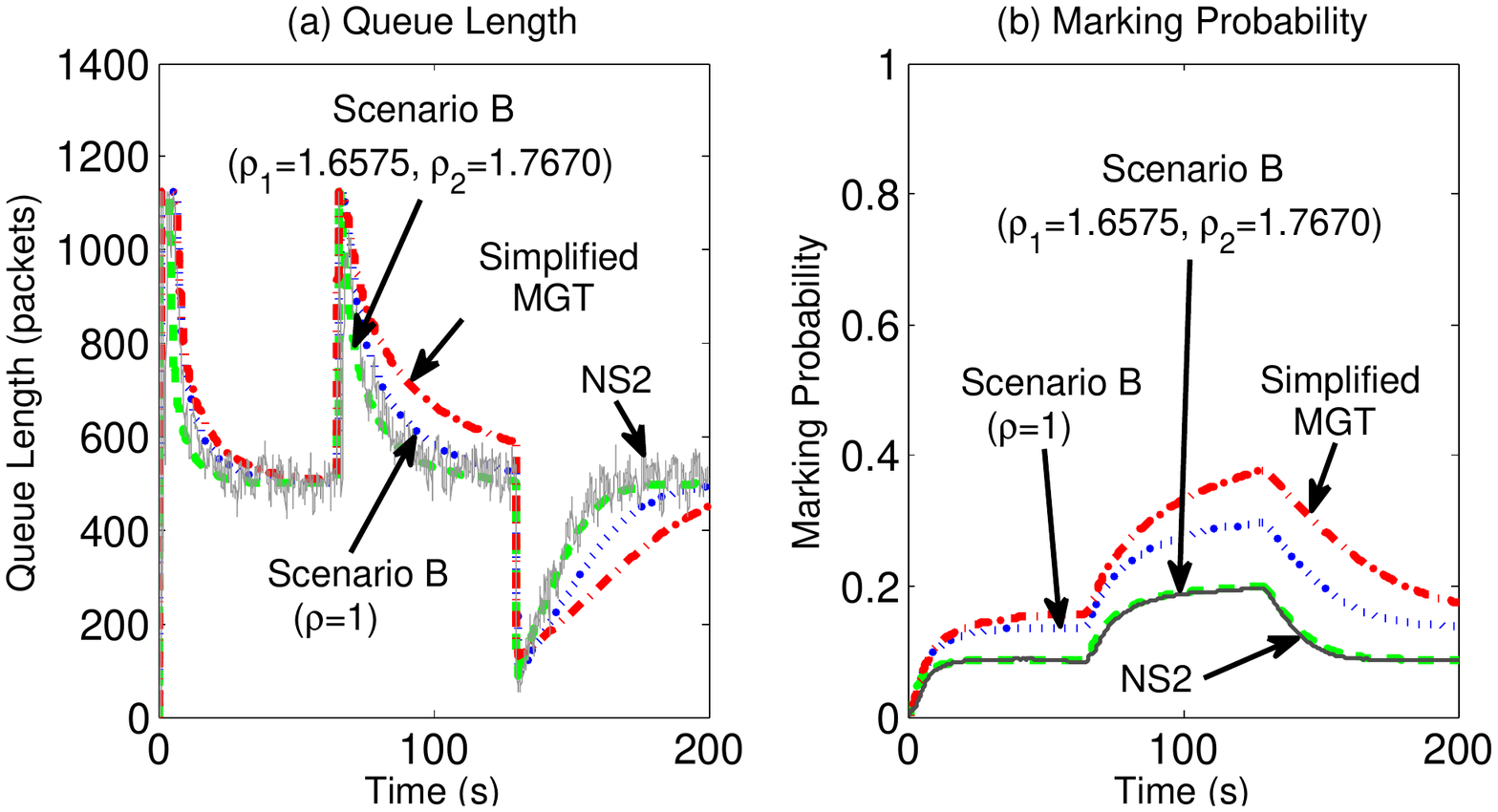}
\caption{Queue length and marking probability versus time for REM AQM when $N$ varies.}
\label{fig:REM300200}
\end{figure}

\begin{figure}[!h]
\centering
\includegraphics[width=0.5\textwidth]{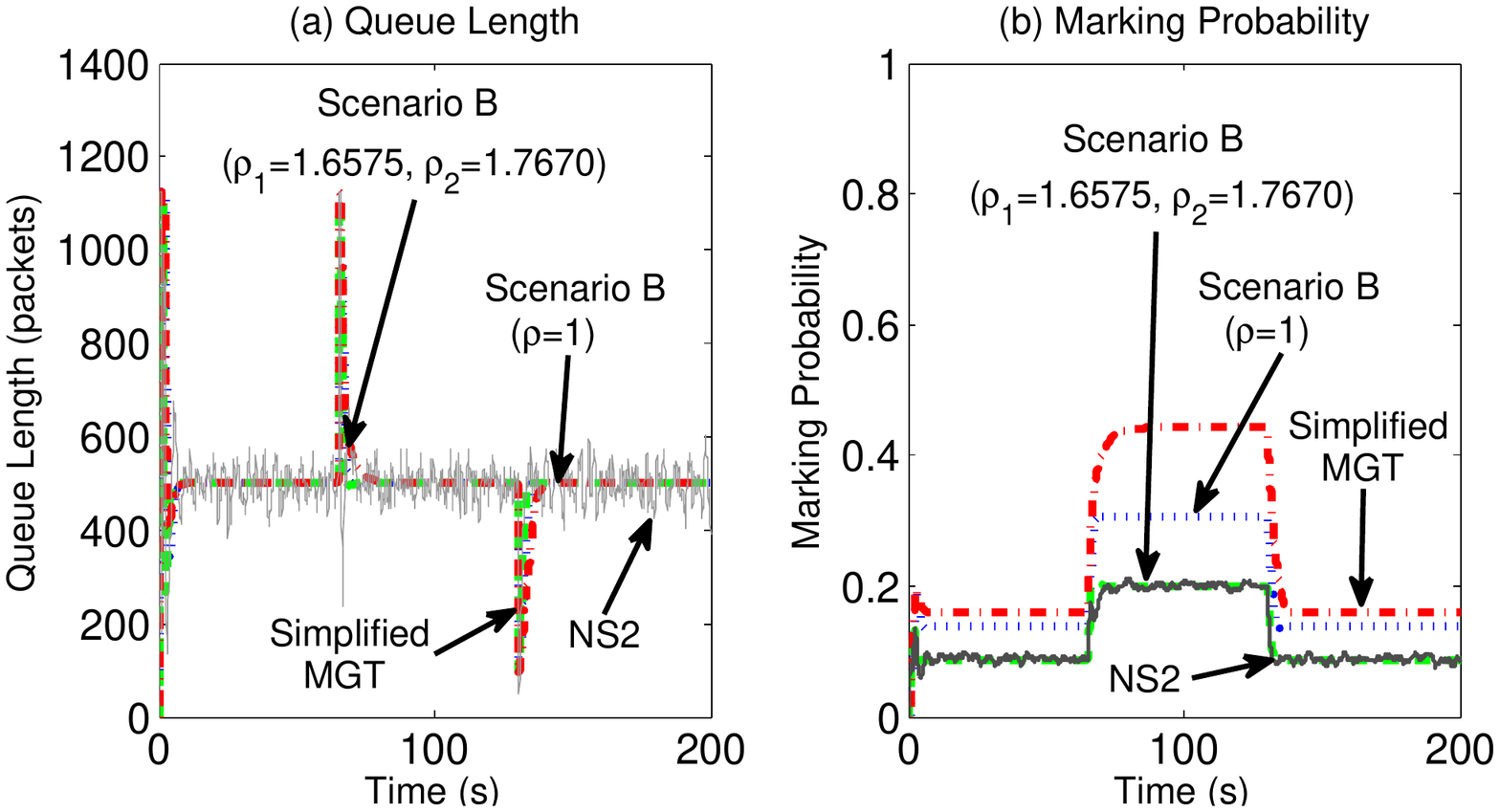}
\caption{Queue length and marking probability versus time for RaQ AQM when $N$ varies.}
\label{fig:RaQ300200}
\end{figure}

The results are shown in Fig.~\ref{fig:PI300200}, Fig.~\ref{fig:REM300200} and  Fig.~\ref{fig:RaQ300200}. In Fig.~\ref{fig:PI300200} and Fig.~\ref{fig:REM300200}, the marking probability of PI and REM AQM schemes have not reach the theoretical values presented in the above table. This is because the TCP/AQM system has not reached steady state when $N$ changes at times 65 s and 130 s. Nevertheless, the results in Fig.~\ref{fig:RaQ300200} confirm the theoretical values due to the fast convergence of the RaQ scheme. As expected, in such Mild and Mild/Moderate congestion region when $N$ changes, the curves of Scenario B ($\rho=1$) are closer to the NS2 simulation curves in both queue dynamic and marking probability compared to the curves of the Simplified MGT model.

\subsection{Different RTT}
Now simulation and analytical results of two different RTTs are presented. The propagation time are set as 0.05 s and 0.15 s, then the corresponding RTTs (\ref{eq:RTT_op}) are $0.05+500/5625=0.1389$ and $0.15+500/5625=0.2389$, respectively. The marking probabilities are 0.2894 and 0.1402 from NS2 simulation, then the corresponding $\rho=2.0107$ and $\rho=1.6984$ can be derived. By (\ref{eq:MGT_p0}) and (\ref{eq:p0b}), the $p_0$ of Simplified MGT model and of Scenario B ($\rho=1$) under different RTTs are provided in Table~\ref{table:p0_RTT}.

\begin{table}[!h]
\centering
\caption{ Values of $p_0$ for Different RTTs}
\label{table:p0_RTT}
\begin{tabular}{|c|c|c|c|}
\hline
RTTs (s) & \tabincell{c}{Simplified \\ MGT} & \tabincell{c}{Scenario B \\ ($\rho=1$)} & NS2 \\
\hline
0.1389 & 0.8191 & 0.4503 & 0.2894 \\
\hline
0.2389 & 0.2769 & 0.2168 & 0.1402 \\
\hline
\end{tabular}
\end{table}

\begin{figure}[!h]
\centering
\includegraphics[width=0.5\textwidth]{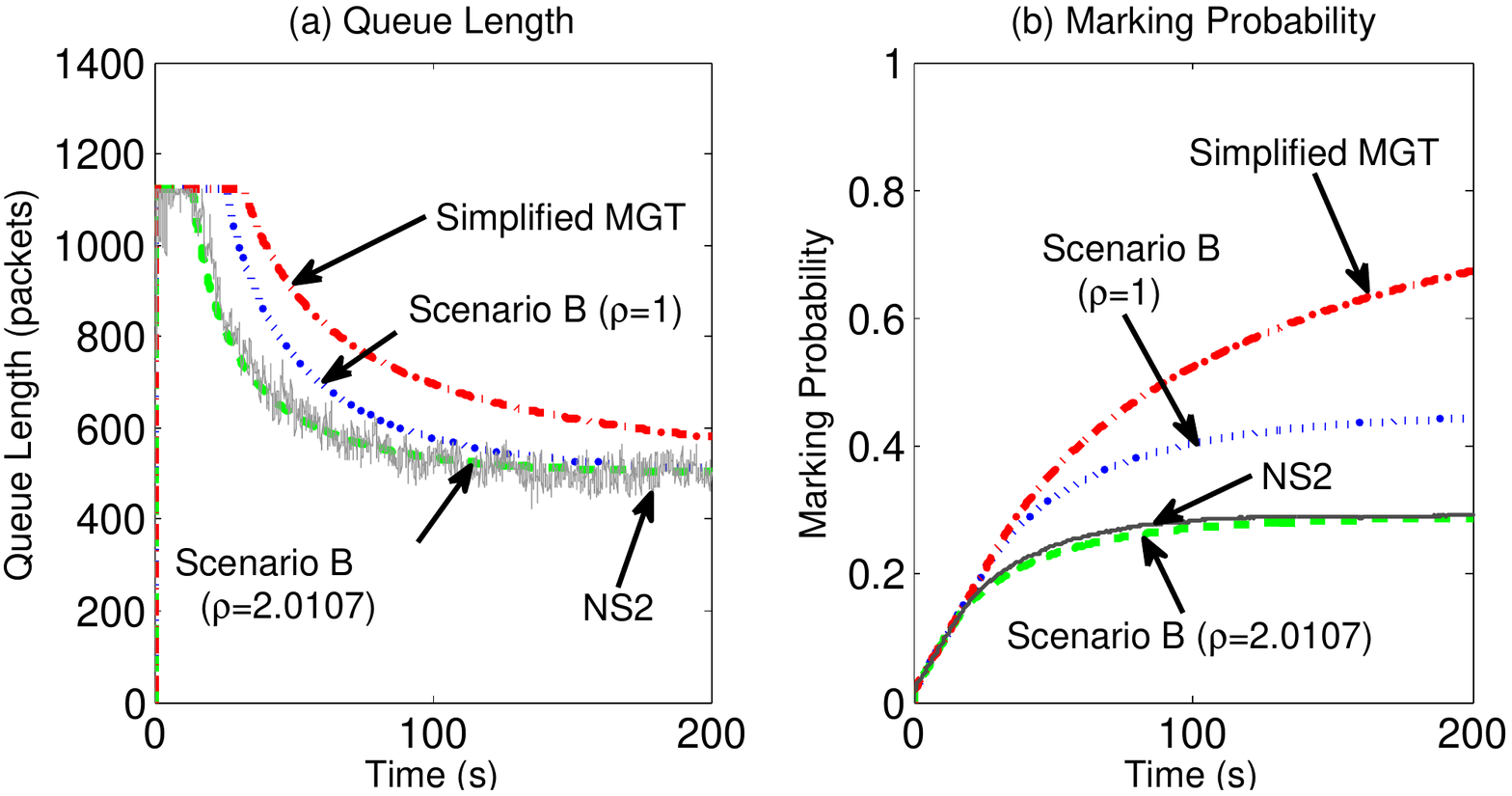}
\caption{Queue length and marking probability versus time for PI AQM with propagation time is 0.05 s.}
\label{fig:PIR005}
\end{figure}

\begin{figure}[!h]
\centering
\includegraphics[width=0.5\textwidth]{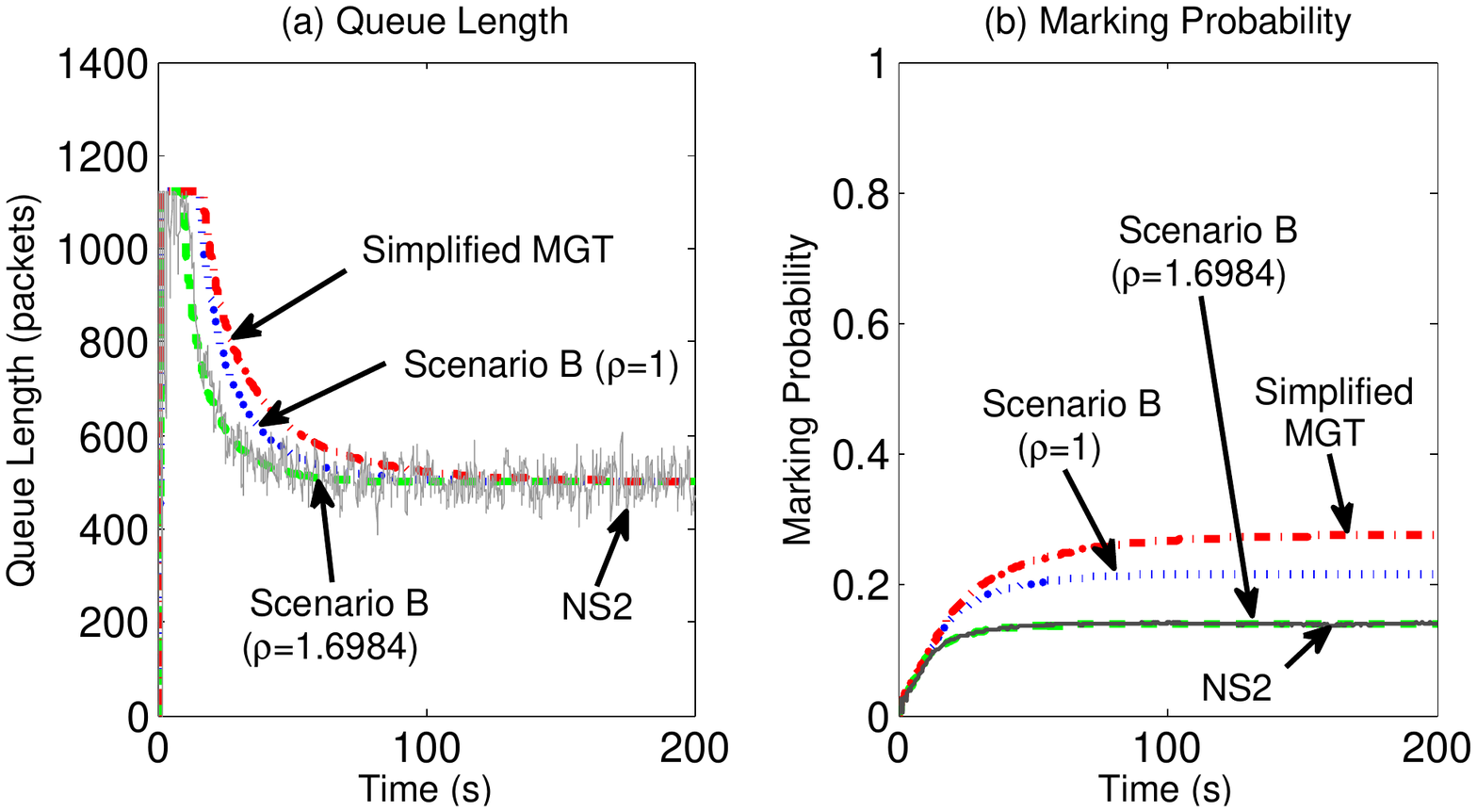}
\caption{Queue length and marking probability versus time for PI AQM with propagation time is 0.15 s.}
\label{fig:PIR015}
\end{figure}

The simulation results are presented in Fig.~\ref{fig:PIR005} and Fig.~\ref{fig:PIR015}, it is shown that Scenario B ($\rho=1$) improves the accuracy of queue length and marking probability in both cases relative to the Simplified MGT model. Note that Scenario B ($\rho$=1.6984) in Fig.\ref{fig:PIR015} matches NS2 simulations better than Scenario B ($\rho=2.0107$) in Fig.\ref{fig:PIR005}. This can be explained by the congestion level. According to (\ref{eq:w0}) and $\overline{w} = W_{s0}/N$, we obtain $\overline{w}=1.5626 $  and $\overline{w}=2.6876$ for $RTT=0.1389$ and $RTT=0.2389$, respectively. As shown in Table~\ref{table:congestion level}, the second case falls into the mildly congested region while the first one falls into the moderately congested region, hence the Scenario B model performs better in the second case.

\subsection{Different Link Capacities}
\begin{figure}[!h]
\centering
\includegraphics[width=0.5\textwidth]{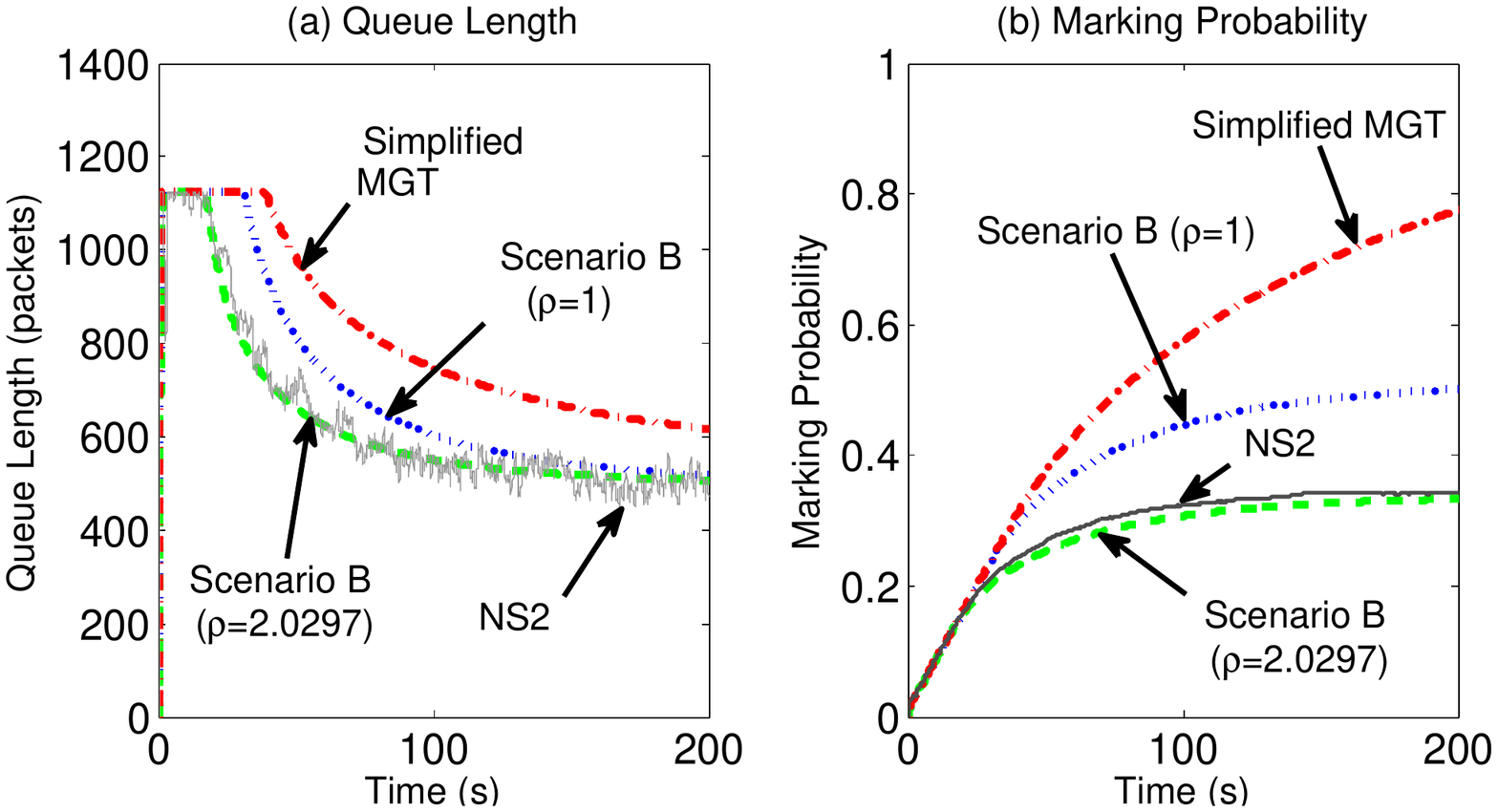}
\caption{Queue length and marking probability versus time for PI AQM with link capacity C is 15 Mb/s.}
\label{fig:PIC15M}
\end{figure}

\begin{figure}[!h]
\centering
\includegraphics[width=0.5\textwidth]{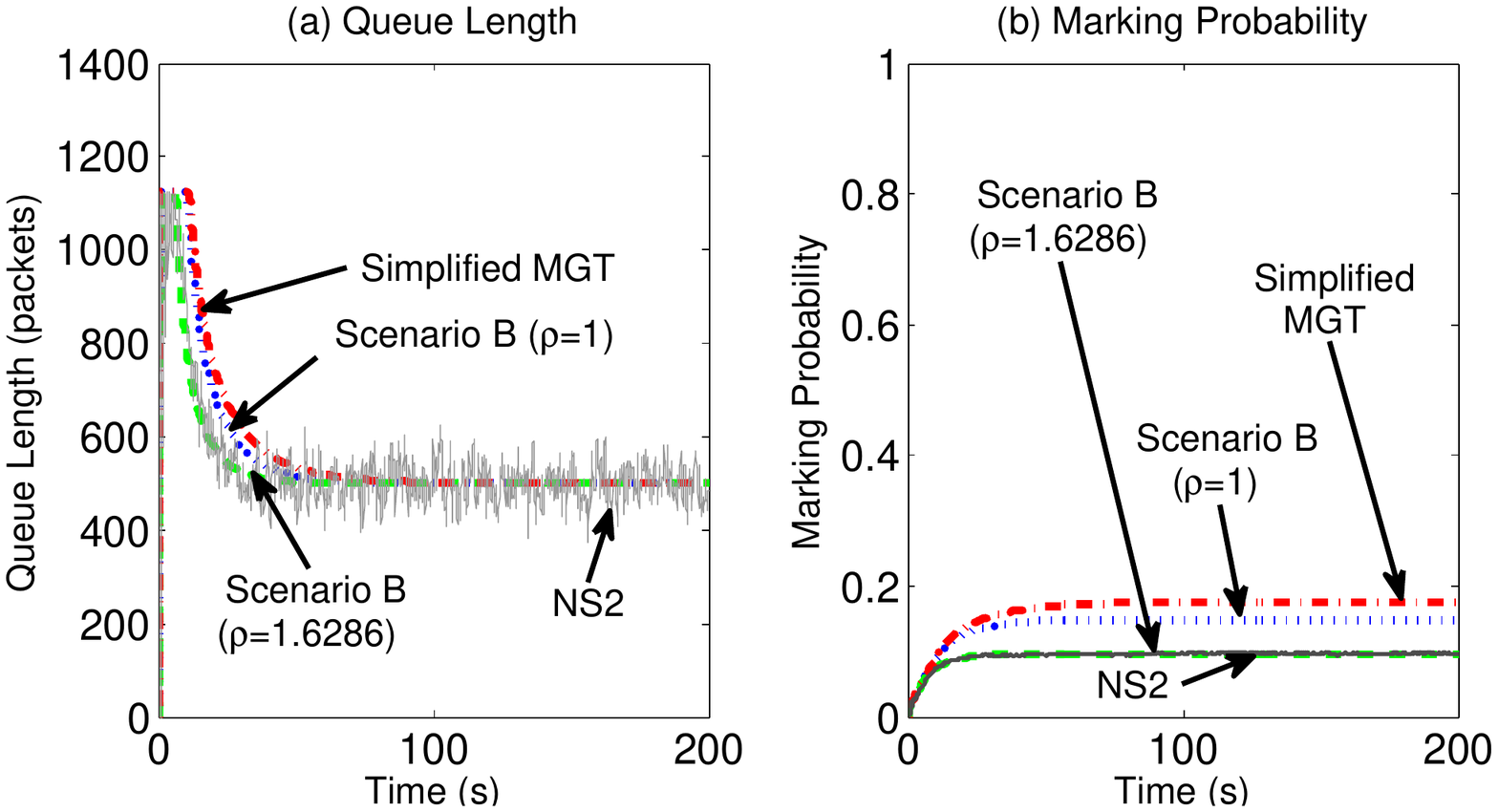}
\caption{Queue length and marking probability versus time for PI AQM with link capacity C is 95 Mb/s.}
\label{fig:PIC95M}
\end{figure}

We now consider two cases where the link capacities are 15Mb/s and 95 Mb/s. By (\ref{eq:MGT_p0}) and (\ref{eq:p0b}), the $p_0$ of Simplified MGT and of Scenario B model ($\rho=1$) for the two cases are given in Table~\ref{table:p0_Link}. From the NS2 simulations, the marking probabilities for the cases of $C=15$ Mb/s and $C=95$ Mb/s are 0.3426 and 0.0973, respectively. Then the corresponding $\rho$ values are 2.0297 and 1.6286, respectively. The results are shown in Fig.~\ref{fig:PIC15M} and Fig.~\ref{fig:PIC95M}. In the two figures, the Scenario B curves with the $\rho$ values of 2.0297 and 1.6286 match the NS2 simulation resuls very well, and the curves generated by Scenario B ($\rho=1$) are closer to the NS2 simulation results than the curves of the Simplified MGT model.

\begin{table}[!h]
\centering
\caption{ Values of $p_0$ for Different Link Capacities}
\label{table:p0_Link}
\begin{tabular}{|c|c|c|c|}
\hline
\tabincell{c}{Link \\ Capacities (Mb)} & \tabincell{c}{Simplified \\ MGT} & \tabincell{c}{Scenario B \\ ($\rho=1$)} & NS2 \\
\hline
15 & 1.0577 & 0.5140 & 0.3426 \\
\hline
95 & 0.1756 & 0.1494 & 0.0973 \\
\hline
\end{tabular}
\end{table}

\subsection{Time Interval $\Delta t$}
\label{subsec:step time}
\begin{figure}[!h]
\centering
\includegraphics[width=0.5\textwidth]{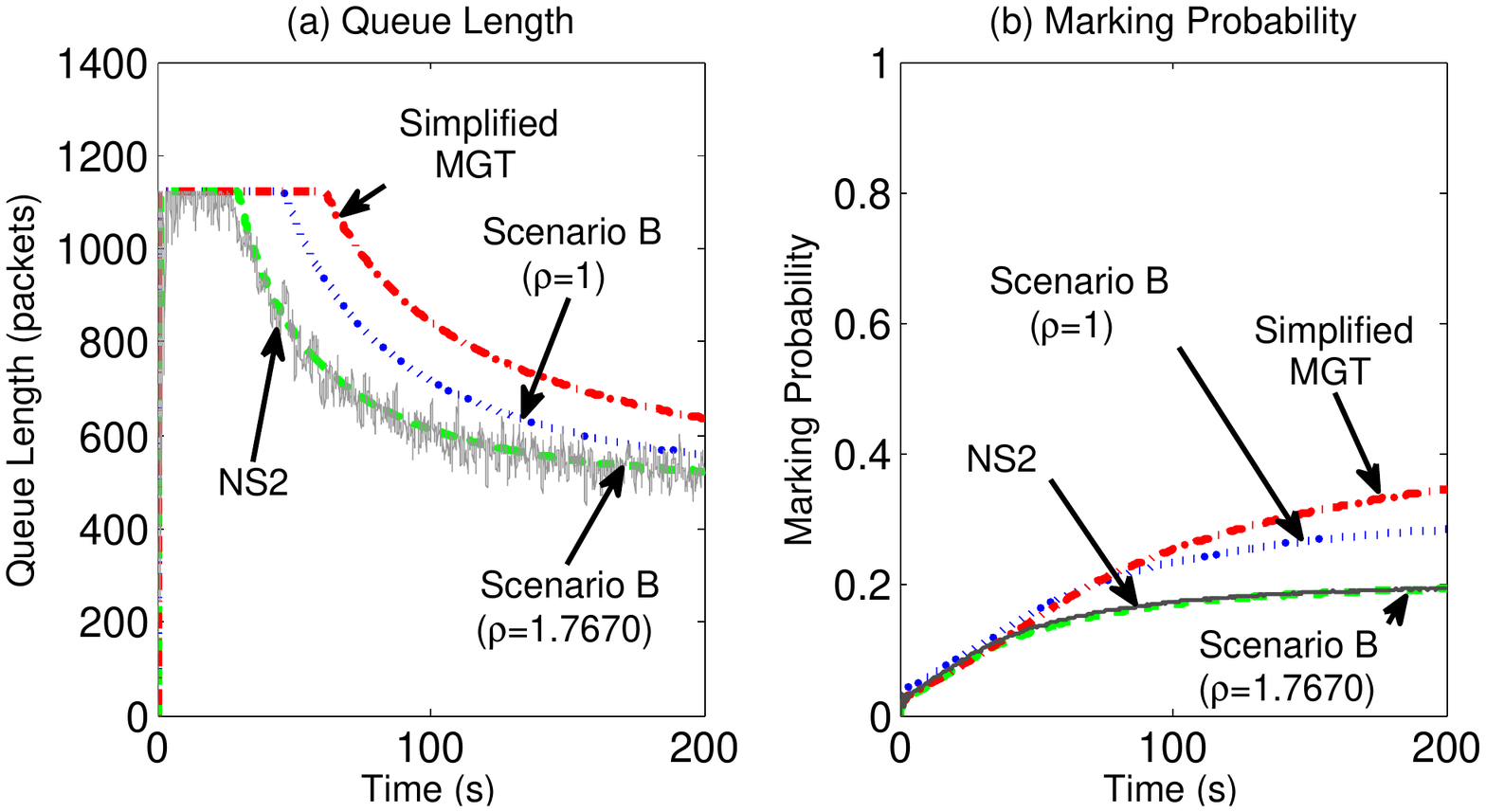}
\caption{Queue length and marking probability versus time for RaQ AQM with step time $\Delta t$ is 0.2 s.}
\label{fig:RaQ02}
\end{figure}

In practice the marking probability may change during the time interval $\Delta t$, while we assume that it stays constant in our model (and also the discrete-time implementation of the Simplified MGT model). Therefore, the longer $\Delta t$ is, the larger the error it introduces. It is required that $\Delta t$ is bounded above by the AQM sampling period denoted by $T$, to allow sufficient time to update the marking probability before the next sampling period. Here we increase the value of $\Delta t$ from its previous value of 0.0005 s to 0.2 s and the AQM sampling period from 0.005 s to 0.2 s, and we examine the effect on the accuracy of our model and of the Simplified MGT model. Notice that the $\Delta t$ and AQM sampling period are now longer that the RTT  which is 0.1889 s by (\ref{eq:RTT_op}). Because of RaQ¡¯s short convergence time, we only present results here for RaQ AQM. The parameter $\rho = 1.7670$ by (\ref{eq:p0b}) is used, and Scenario B is chosen of this case. The result is shown in Fig.~\ref{fig:RaQ02}, which demonstrates that even if $\Delta t$ is greater than the RTT, the Scenario B model (with $\rho=1.7670$) still has the capability to track the simulation results well. The performance of the Simplified MGT model is also quite good under the same settings, while it is not as good as that of Scenario B (with $\rho=1$) for both the queue length and marking probability dynamics.

\section{CONCLUSIONS}
\label{sec:Conclusions}

In this paper, we have proposed a new model for a time-driven TCP/AQM system. This new model rigorously considers the four algorithms embedded in TCP, and it contributes to performance studies of AQM schemes. We have provided both discrete and continuous-time models. Our proposed model is based on bounds associated with heavy and light congestion, but we have shown that the bounds are tight in steady state and the model is applicable to a wide range of congestion levels. In addition, the model has been validated using extensive simulation results for a wide range of parameters and various AQM schemes and it was demonstrated that it is more accurate than the Simplified MGT model. In fact, the accuracy of our model was consistently demonstrated in all cases studied.

\appendix[Stability under Linearization ]
\label{sec:app1}
Recalling Scenario A described by (\ref{eq:SceAcon}), we define:
\begin{align}
f(W_s, W_{sR}, p_R, q_R)\doteq & \dot{W_s}(t) = \frac{W_{sR}}{T_p + \frac{q_R}{C}} (1-p_R)  \nonumber \\
&-\frac{\rho W_s W_{sR} }{2N(T_p + \frac{q_R}{C})} p_R ,
\end{align}

where $W_{sR}(t) \doteq W_s(t-R)$, $p_R(t) \doteq p(t-R)$, $q_R(t) \doteq q(t-R)$.

Taking partial derivatives at the operating point ($W_{s0}, q_0, p_0$) of this model yields:
\begin{equation}
\frac{\partial f}{\partial W_s}=-\frac{\rho W_{s0} p_0}{2NR_0}
\end{equation}

\begin{equation}
\frac{\partial f}{\partial W_{sR}} = \frac{1-p_0}{R_0} - \frac{\rho W_{s0} p_0}{2NR_0}
\end{equation}

\begin{equation}
\frac{\partial f}{\partial p_R} = -\frac{W_{s0}}{R_0} - \frac{\rho W_{s0}^2}{2NR_0}
\end{equation}

\begin{equation}
\frac{\partial f}{\partial q_R} = -\frac{W_{s0}(1-p_0)}{R_0^2 C} + \frac{\rho W_{s0}^2 p_0}{2NR_0^2 C} .
\end{equation}

Hence Scenario A linearized form is expressed as:
\begin{align}
\delta\dot{W_s}(t)= &\frac{\partial f}{\partial W_s} \delta W_s(t) + \frac{\partial f}{\partial W_{sR}} \delta W_s(t-R) \nonumber \\
                    & + \frac{\partial f}{\partial p_R} \delta p(t-R) + \frac{\partial f}{\partial q_R} \delta q(t-R) ,
\label{eq:linearized model}
\end{align}
where
\begin{equation}
 \begin{cases}
\ \delta W_s(t) \doteq W_s(t) - W_{s0}\\
\ \delta q(t) \doteq q(t)  - q_0\\
\ \delta p(t) \doteq p(t) - p_0.
 \end{cases}
\end{equation}

We linearize Scenario B model in a similar way to (\ref{eq:linearized model}), where the partial derivatives are replaced by:
\begin{equation}
 \begin{cases}
\dfrac{\partial f}{\partial W_s} = -\dfrac{N(1-p_0)}{R_0W_{s0}} - \dfrac{\rho W_{s0} p_0}{2NR_0}\\
\dfrac{\partial f}{\partial W_{sR}} = \dfrac{N(1-p_0)}{R_0W_{s0}} - \dfrac{\rho W_{s0} p_0}{2NR_0}\\
\dfrac{\partial f}{\partial p_R} = -\dfrac{N}{R_0} - \dfrac{\rho W_{s0}^2}{2NR_0}\\
\dfrac{\partial f}{\partial q_R} = -\dfrac{N(1-p_0)}{R_0^2 C} + \dfrac{\rho W_{s0}^2 p_0}{2NR_0^2 C} .
 \end{cases}
 \label{eq:parameters of linearized B}
\end{equation}

Linearizing (\ref{eq:DeltaQ_Con_ECN}) when ENC is enabled gives:
\begin{equation}
\delta\dot{q}(t)=\frac{1}{R_0} \delta W_s(t) - \frac{W_{s0}}{R_0^2 C}\delta q(t) .
\end{equation}

\begin{figure}[!h]
\centering
\includegraphics[width=0.5\textwidth]{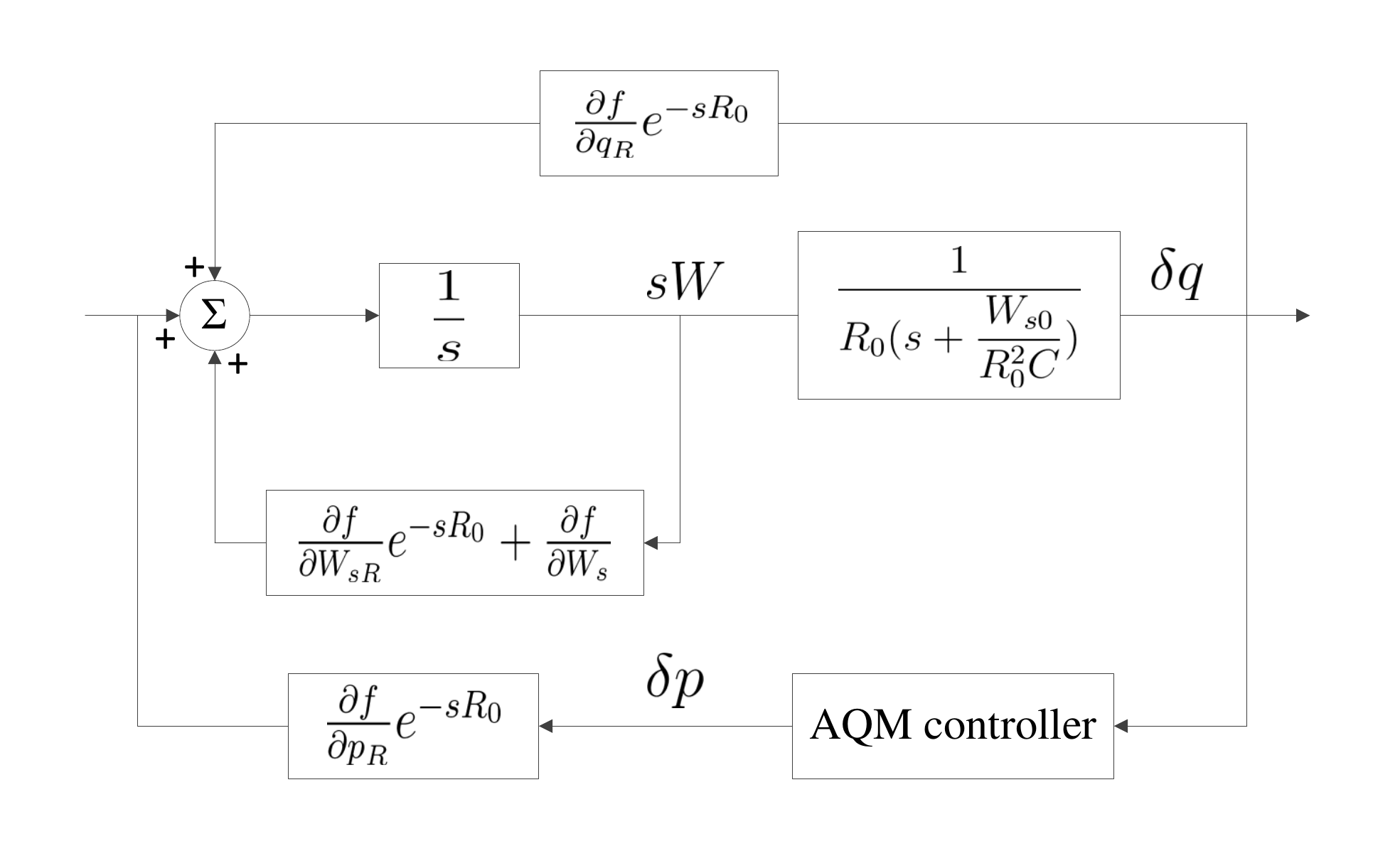}
\caption{The block diagram of TCP/AQM system.}
\label{fig:block1}
\end{figure}

\begin{figure}[!h]
\centering
\includegraphics[width=0.5\textwidth]{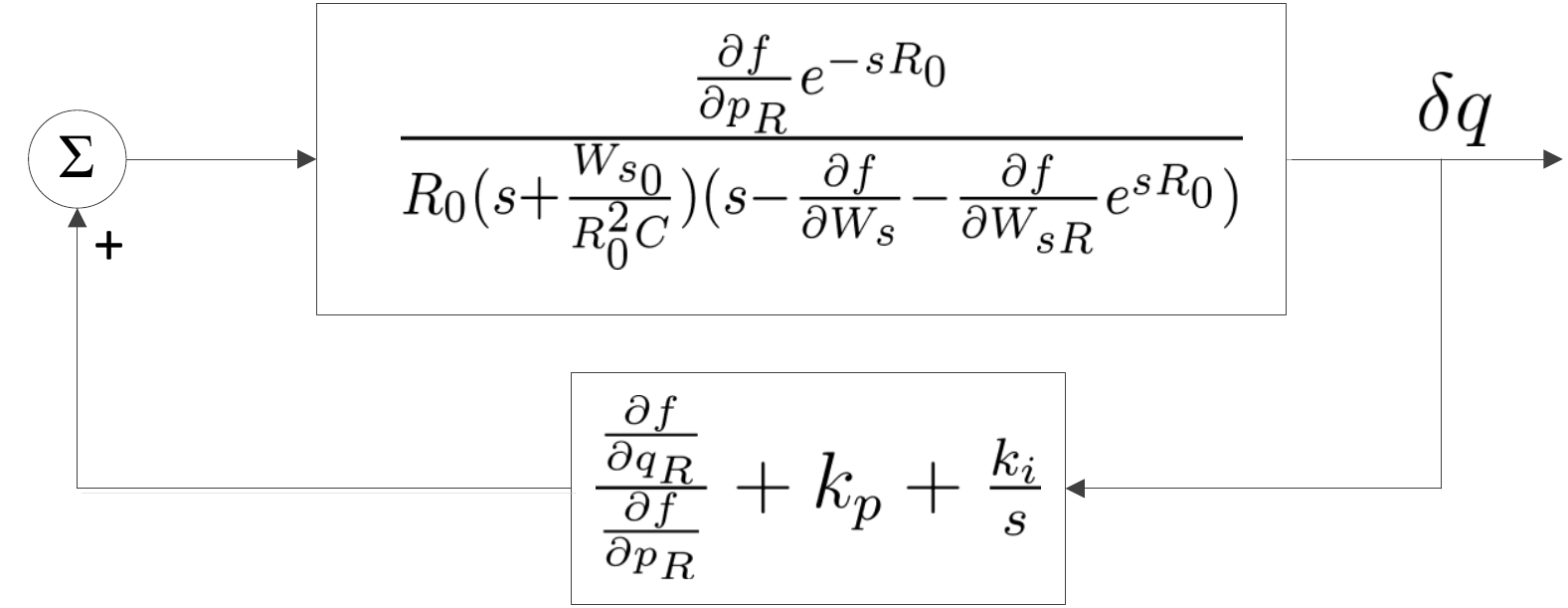}
\caption{Simplified block diagram of TCP/AQM system.}
\label{fig:block2}
\end{figure}

We obtain the block diagram presented in Fig.~\ref{fig:block1}. We choose PI as the AQM algorithm and we obtain a closed-loop system. In Laplace area, the PI controller has the form $k_p + \frac{k_i}{s}$, where $k_p$ and $k_i$ are the proportion and integral coefficient, respectively. Then the diagram of Fig.~\ref{fig:block1} is transformed to the diagram of Fig.~\ref{fig:block2}. For simplicity, we approximate the time delay as the first-order lag \cite{2006:Franklin} $e^{sR_0} \approx \frac{1}{1+sR_0}$. Thus, we obtain the characteristic equation:
\begin{equation}
 s^4 + \alpha_1 s^3 + \alpha_2 s^2 + \alpha_3 s + \alpha_4=0 ,
\end{equation}
where
\begin{equation}
 \begin{cases}
\ \alpha_1 \doteq \frac{1}{R_0} - \dfrac{\partial f}{\partial W_{s}} + \dfrac{W_{s0}}{R_0^2 C}\\
\ \alpha_2 \doteq -\dfrac{\dfrac{\partial f}{\partial W_{s}}+\dfrac{\partial f}{\partial W_{sR}}}{R_0} + \dfrac{W_{s0}(1-\dfrac{\partial f}{\partial W_{s}} R_0)}{R_0^3 C}\\
\ \alpha_3 \doteq -\dfrac{W_{s0}(\dfrac{\partial f}{\partial W_{s}}+\dfrac{\partial f}{\partial W_{sR}})}{R_0^3 C} - \dfrac{\dfrac{\partial f}{\partial q_R} + \dfrac{\partial f}{\partial p_R} k_p}{R_0^2}\\
\ \alpha_4 \doteq -\dfrac{\dfrac{\partial f}{\partial p_R} k_i}{R_0^2} .
 \end{cases}
\end{equation}

According to the Routh stability criterion, the system is stable if and only if
\begin{equation}
 \begin{cases}
\ \alpha_1 >0\\
\ \beta_1 \doteq \alpha_1 \alpha_2 - \alpha_3 >0\\
\ \beta_2 \doteq \alpha_3(\alpha_1 \alpha_2 - \alpha_3) - \alpha_1^2 \alpha_4>0\\
\ \alpha_4 >0 .
 \end{cases}
\end{equation}

As an example we demonstrate that the settings of the Subsection~\ref{sub:Perfo_ConLev} leads to a stable system. Computation based on these settings are given in Table~\ref{tab:stability A} and \ref{tab:stability B}, where PI is chosen as the AQM scheme. The values shown in the two tables demonstrate that they all satisfy the conditions, so the system is stable.

\begin{table}[!h]
\centering
\caption{The Values for Scenario A }
\label{tab:stability A}
\begin{tabular}{|c|c|c|c|c|}
\hline
\diagbox{$N$}{Values} & $\alpha_1$ & $\beta_1$ & $\beta_2$ & $\alpha_4$ \\
\hline
500 & 14.8205 & 946.6351 & 1.2581e+3 & 0.0472 \\
\hline
800 & 14.0721 & 799.4057 & 8.3581e+4 & 0.0270 \\
\hline
1100 & 13.6513 & 732.6899 & 6.7882e+4 & 0.0225 \\
\hline
2000 & 13.2988 & 672.6120 & 5.5029e+4 & 0.0194 \\
\hline
\end{tabular}
\end{table}

\begin{table}[!h]
\centering
\caption{The Values for Scenario B }
\label{tab:stability B}
\begin{tabular}{|c|c|c|c|c|}
\hline
\diagbox{$N$}{Values} & $\alpha_1$ & $\beta_1$ & $\beta_2$ & $\alpha_4$ \\
\hline
200 & 12.4921 & 536.3968 & 3.5160e+4 & 0.0403 \\
\hline
500 & 45.8829 & 2.6924e+4 & 2.6975e+7 & 0.0422 \\
\hline
800 & 15.7664 & 1.1551e+3 & 1.7474e+5 & 0.0203 \\
\hline
1100 & 16.9312 & 1.4268e+3 & 2.6368e+5 & 0.0232 \\
\hline
\end{tabular}
\end{table}

\section*{ACKNOWLEDGEMENT}
This paper was supported by a grant from City University of Hong Kong (Project No. 9380044), and two grants from National Natural Science Foundation of China (Project No. 60974129) and (Project No. 70931002).

\bibliographystyle{IEEEtran}
\bibliography{xu_fan}

\vspace*{-2\baselineskip}
\begin{IEEEbiography}[{\includegraphics[width=1in,height=1.25in,clip,keepaspectratio]{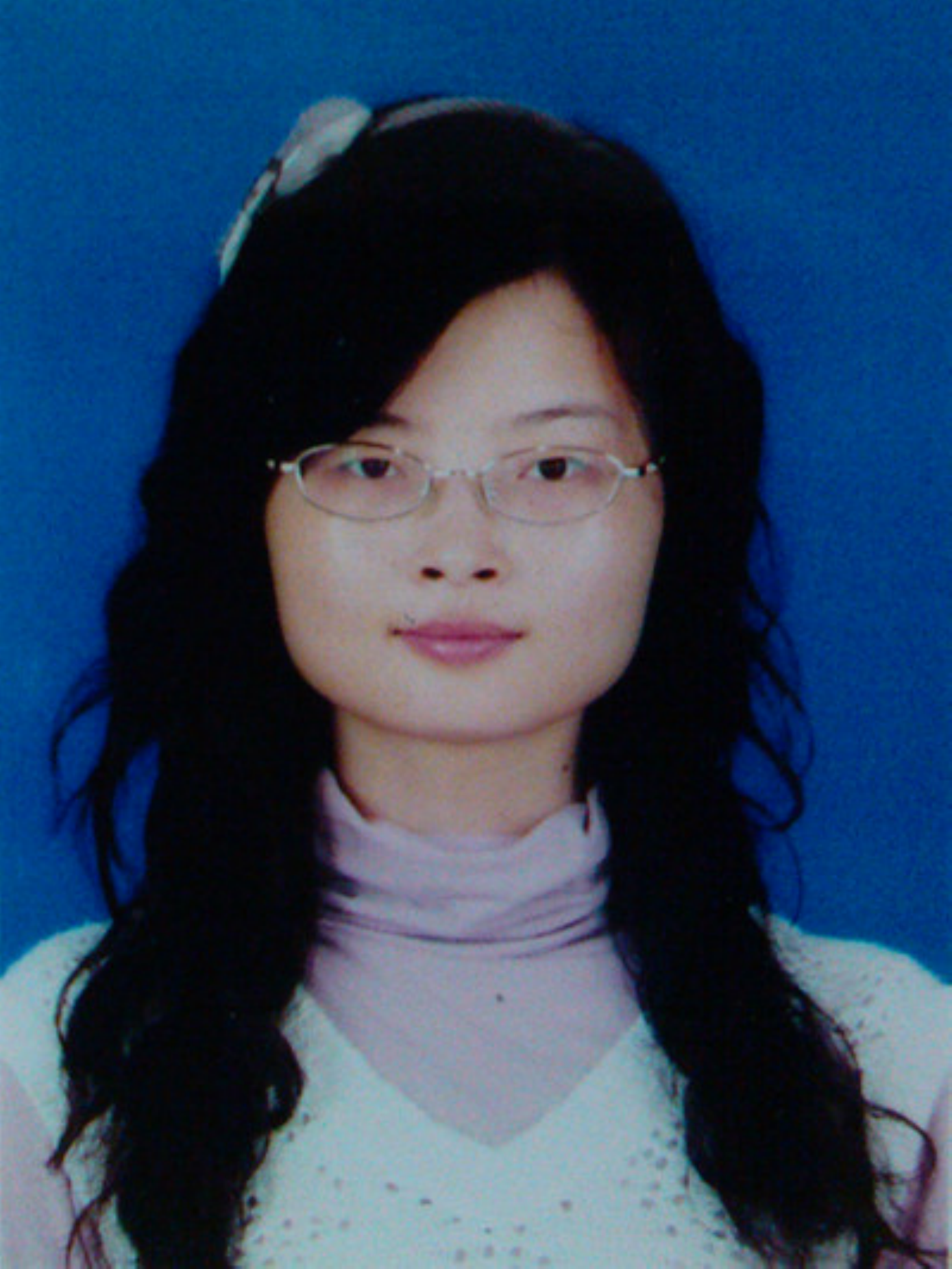}}]{Qin Xu}
is a PhD candidate in the School of Automation, Nanjing University of Science and Technology, Nanjing, P. R. China. She received her B. Eng degree in electronic information engineering from the same school. Her research interest is network congestion control.
\end{IEEEbiography}

\vspace*{-2\baselineskip}
\begin{IEEEbiography}[{\includegraphics[width=1in,height=1.25in,clip,keepaspectratio]{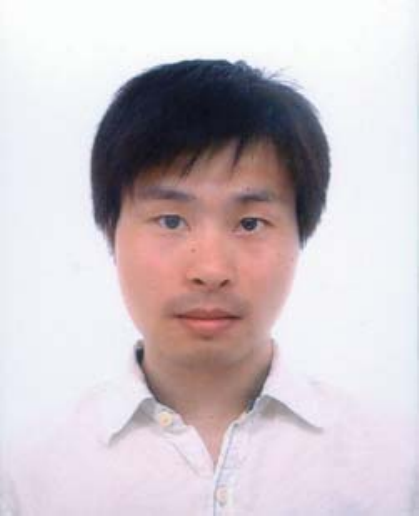}}]{Fan Li}
is a research assistant with the department of Electronic Engineering, City University of Hong Kong. He received his MSc. degree from the same department in 2009, and the B.Eng. degree in electronic information engineering from Nanjing University of Science and Technology, Nanjing, P. R. China. His research interests include teletraffic theory, network transmission control, and optical network dimensioning.
\end{IEEEbiography}

\vspace*{-2\baselineskip}
\begin{IEEEbiography}[{\includegraphics[width=1in,height=1.25in,clip,keepaspectratio]{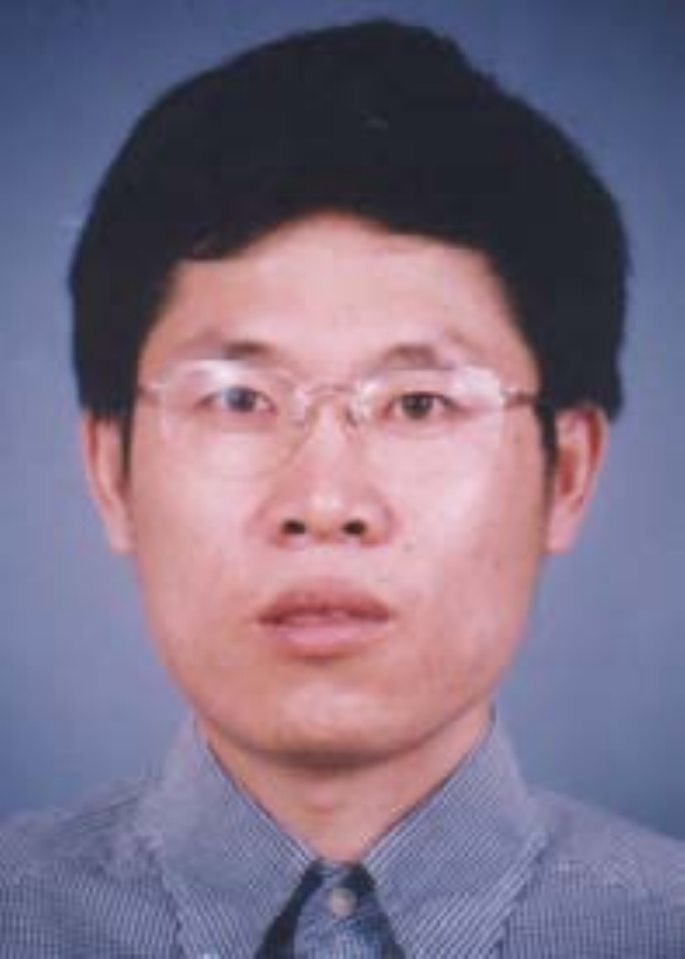}}]{Jinsheng Sun}received the B.S., M.S. and Ph.D. degrees in Control Science from Nanjing University of Science and Technology in 1990, 1992 and 1995, respectively. Since 1995, he has been with the Department of Automation, Nanjing University of Science and Technology. Currently, he is a full professor. In 2006 and 2007, he took up a Research Fellow position at the Department of Electrical and Electronic Engineering, The University of Melbourne, Victoria, Australia. His research interests include congestion control and fault-tolerant control.
\end{IEEEbiography}

\vspace*{-2\baselineskip}
\begin{IEEEbiography}[{\includegraphics[width=1in,height=1.25in,clip,keepaspectratio]{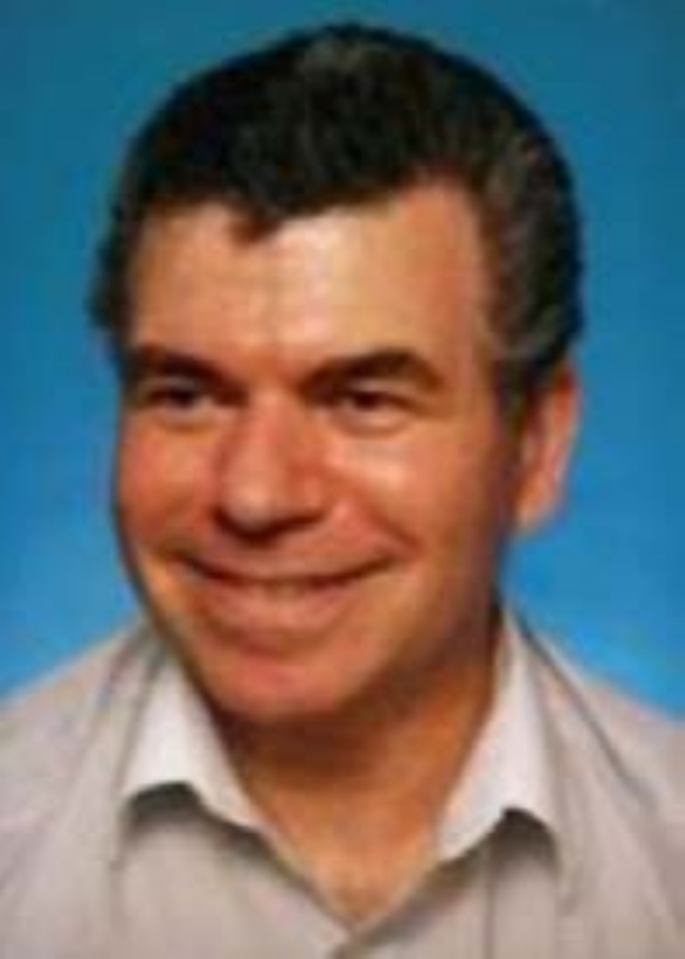}}]{Moshe Zukerman}
(M'87-SM'91-F'07) received his B.Sc. and M.Sc. degrees from the Technion, and his Ph.D. degree from UCLA in 1985. During 1986-1997, he was with the Telstra Research Laboratories, first as a Research Engineer and, in 1988-1997, as a Project Leader. During 1997-2008, he was with The University of Melbourne, Victoria, Australia. In 2008 he joined City University of Hong Kong as a Chair Professor of Information Engineering, and a Team Leader. He has over 250 publications in scientific journals and conference proceedings. He has served on various editorial boards and technical program committees.
\end{IEEEbiography}
\end{document}